\documentclass[
    aip,
    amsmath,amssymb,
    reprint
]{revtex4-1}

\usepackage[utf8]{inputenc}
\usepackage[T1]{fontenc}
\usepackage{lipsum}  
\usepackage{makecell}
\usepackage{mathtools}
\usepackage{amsmath}
\usepackage{setspace}
\usepackage{amssymb}
\usepackage{enumitem}
\setlist{nosep}
\usepackage{algpseudocode}
\usepackage[dvipsnames]{xcolor}
\usepackage{graphicx}
\usepackage{bm}
\usepackage{mathtools}
\usepackage[dvipsnames]{xcolor}
\usepackage{etoolbox}
\usepackage[hidelinks]{hyperref}

\newcommand{\MM}{\textit{Materials and Methods}}
\renewcommand{\descriptionlabel}[1]{%
    \hspace{\labelsep}\normalfont\bfseries #1\unskip
}
\setlist[description]{
    leftmargin=0pt,
    itemindent=0pt,
    labelsep=0pt
}

\setlist[enumerate]{
    leftmargin=2em,
    labelsep=0.4em
}
\makeatletter
\def\@email#1#2{%
  \endgroup
  \patchcmd{\titleblock@produce}
    {\frontmatter@RRAPformat}
    {\frontmatter@RRAPformat{%
       \produce@RRAP{*#1\href{mailto:#2}{#2}}%
     }\frontmatter@RRAPformat}
    {}{}
}
\makeatother

\usepackage{hyperref}
\hypersetup{
	colorlinks,
	linkcolor={red!60!black},
	citecolor={red!60!black},
	urlcolor={red!80!black}
}

\begin{document}

\title[Simulation-free approximate Bayesian computation]
{Simulation-free approximate Bayesian computation
for stochastic reaction networks}

\author{Zekai Li}
\author{Mauricio Barahona}
\thanks{To whom correspondence should be addressed. Electronic mail: p.thomas@imperial.ac.uk and m.barahona@imperial.ac.uk}
\author{Philipp Thomas}
\thanks{To whom correspondence should be addressed. Electronic mail: p.thomas@imperial.ac.uk and m.barahona@imperial.ac.uk}
\affiliation{
Department of Mathematics, Imperial College London,
London SW7 2AZ, United Kingdom
}

\date{\today}

\begin{abstract}
Uncovering stochastic reaction networks from data is central to single-cell biology. Simulation-based inference has evolved into a flexible paradigm for this purpose, circumventing the computational intractability of the likelihood. Notably, approximate Bayesian computation performs this inference by carrying out extensive Gillespie simulations, usually at a high computational cost. Here, we propose a simulation-free approximate Bayesian computation (SFABC) framework that avoids both simulation and explicit likelihood solutions of the network dynamics. Instead, SFABC exploits the Chemical Master Equation to formulate a tractable convex optimization with ensuing constraints, which are then used to accept sampled parameters consistent with the observed data. The resulting posterior approximation comes with theoretical guarantees: it can be tightened systematically, and it provably converges to the exact Bayesian posterior distributions. Using synthetic benchmarks, we show that SFABC matches the accuracy of standard simulation-based algorithms at a substantially reduced computational cost. We then showcase the applicability of SFABC to experimental systems biology by analyzing gene transcription bursting kinetics from snapshot, time-resolved, and cell-cycle-resolved single-cell data.
\end{abstract}

\maketitle

\section{Introduction}

With the advent of single-cell sequencing technologies, large-scale transcriptomic datasets have become available, enabling the study of single-cell dynamics and providing insights into complex biological mechanisms spanning gene expression heterogeneity \citep{swain2002intrinsic, suter2011mammalian, dar2012transcriptional, zechner2014scalable, dattani2017stochastic, larsson2019genomic}, molecular interaction networks \citep{tian2006stochastic, thomas2014phenotypic, trevino2021chromatin, kartha2022functional, sun2022scrna}, or differentiation trajectories \citep{ bergen2020generalizing, qiao2021representation, gorin2022rna}. To capture the stochasticity arising from a low number of molecules per cell, these dynamics are often modeled as stochastic reaction networks (SRNs) that can be analyzed using the Chemical Master Equation (CME) and Gillespie's stochastic simulation algorithm \citep{gillespie1977exact}, in which reaction events occur randomly with propensities determined by the network structure and reaction rates \citep{shahrezaei2008analytical, jiang2021neural, wrede2022robust, tang2023neural, liu2024distilling}. Inferring these reaction rate parameters is therefore crucial for understanding the underlying biological mechanisms, but it remains challenging given the stochastic nature of cellular dynamics %and the uncertainty in the data 
\cite{schnoerr2017approximation, warne2019simulation}. 

Numerous methods for parameter inference from single-cell data have been developed over the past decades. Likelihood-based approaches estimate parameters by explicitly modeling the probability of the observed data through maximum likelihood estimation \citep{suter2011mammalian,larsson2019genomic,gorin2025monod} or Bayesian inference \citep{kim2013inferring,jiang2022identification, edwards2023synergising}. However, likelihood evaluation requires analytical or numerical solutions to the CME that present an infinite system of equations and can become intractable even for relatively simple networks \cite{schnoerr2017approximation} and require further approximations \cite{frohlich2016inference,munsky2018distribution,catanach2020bayesian}. Likelihood-free methods offer a tangible solution by avoiding explicit evaluation of the likelihood. Although moment approximations provide a frequentist likelihood-free route \citep{frohlich2016inference, backenkohler2017moment,li2025moment}, most Bayesian likelihood-free approaches still rely on stochastic simulations to generate realizations of stochastic reaction networks \cite{lillacci2013signal, lenive2016inferring,mikelson2020likelihood,tang2023modelling,volteras2024global,pessoa2026simulation}, making simulation-based inference (SBI) a central paradigm in likelihood-free Bayesian inference \citep{cranmer2020frontier}.

A limitation of SBI is the trade-off between estimation accuracy and computational cost, since improving accuracy typically requires more simulations. As a classical SBI approach, approximate Bayesian computation (ABC) estimates the posterior by repeatedly simulating data under parameter proposals and comparing the simulated data with observations \citep{tavare1997inferring, Pritchard1999,toni2009approximate, wilkinson2013approximate, liepe2014framework}. Although ABC has provable convergence to the true posterior, this convergence comes at the cost of rapidly increasing computational burden \citep{blum2010approximate, barber2015rate, frazier2018asymptotic}, despite the development of more efficient sampling schemes and other algorithmic improvements \citep{marjoram2003markov,sisson2007sequential,wegmann2009efficient, toni2010simulation, wood2010statistical, del2012adaptive, park2016k2}. Alternatively, recent advances in machine learning have accelerated SBI through both surrogate models of the simulation process \citep{fearnhead2012constructing, meeds2014gps, wilkinson2014accelerating, jiang2017learning, an2019accelerating, aakesson2021convolutional, frazier2021robust, frazier2023bayesian} and neural posterior estimation methods that learn the posterior directly from simulated data \citep{izbicki2019abc, rodrigues2021hnpe, ward2022robust, tang2023modelling}. However, these neural approximations can introduce inductive bias that forfeits convergence guarantees, and so the quality of SBI must be assessed on a case-by-case basis since it critically depends on the number of simulations available for training \citep{price2018bayesian}.  

In this work, we introduce simulation-free approximate Bayesian computation (SFABC) for SRNs, a novel framework for likelihood-free Bayesian inference that circumvents the need for stochastic simulations. Grounded in the CME, the system of coupled differential equations that governs the dynamics of an SRN, SFABC checks whether a sampled parameter set is compatible with a candidate state probability obtained from the observed data (Fig.~\ref{fig:Fig1}). We prove that the approximated posterior from SFABC converges to the true posterior, subject to several limits that involve zero tolerance. Using synthetic datasets, we demonstrate that SFABC is compatible with various sampling schemes and achieves accurate posterior estimates, while substantially reducing computational cost as the number of observed cells increases. Applied to snapshot scRNA-seq data, SFABC uncovers differences in transcriptional kinetics between genes with different promoter architectures and reveals their posterior dependence structure. With temporal data, SFABC enables inference of kinetic rates on an absolute time scale and recovers associations across data modalities. Finally, we apply SFABC to distinguish among competing models of cell cycle-dependent regulation.

\section{Results}
We consider Bayesian inference of rate parameters $\boldsymbol{k}\coloneqq(k_1,\ldots,k_R)^\intercal\in \mathbb{R}_+^R$ in an SRN with molecular species $\boldsymbol{Y}\coloneqq(Y_s)_{s\in\mathcal{S}'}$, where $\mathcal{S}'$ is the species index set, participating in $r = 1,\ldots,R$ reactions
\begin{equation} \label{eqn:SRN}
\quad \sum_{s \in \mathcal{S}'} v^{-}_{s r}Y_s \xrightarrow{k_ra_r} \sum_{s \in \mathcal{S}'} v^{+}_{s r}Y_s.
\end{equation} 
In reaction $r$, $v^{\pm}_{sr}$ molecules of species $Y_s$ are consumed and produced, respectively, and $a_r$ is the state-dependent, unit-rate propensity function. Let us consider a dataset $X\coloneqq (\boldsymbol{x}_{1}, \ldots,\boldsymbol{x}_{N})\in \mathbb{N}^{|\mathcal{S}|\times N}$ that contains $N$ independent observations of counts of the observable species $\mathcal{S}\subseteq\mathcal{S}'$ in cells at time points $\boldsymbol{\tau} \coloneqq(\tau_1,\ldots,\tau_N)^\intercal\in\mathbb{R}_+^N$. Then the posterior distribution of the rate parameters $\boldsymbol{k}$ is:
\begin{equation} \label{eqn:posterior}
\underbrace{P(\boldsymbol{k} | X)}_{\text{posterior}} \propto  \underbrace{\prod_{i=1}^N \boldsymbol{p}_{\boldsymbol{k},\tau_i}(\boldsymbol{x}_i)}_{\text{likelihood}}\ \underbrace{\pi(\boldsymbol{k})}_{\text{prior}},
\end{equation}
where $\boldsymbol{p}_{\boldsymbol{k},t}(\boldsymbol{x})$ denotes the marginalized probability over the observed species, and the full state probability $\boldsymbol{p}_{\boldsymbol{k},t}$ satisfies the CME
\begin{equation} \label{eqn:CME}
    \frac{\mathrm{d} \boldsymbol{p}_{\boldsymbol{k},t}}{\mathrm{d} t} = \sum_{r=1}^R k_r Q_r \boldsymbol{p}_{\boldsymbol{k},t} \, .
\end{equation} 
Here, the unit-rate transition matrix is defined by $Q_r(\boldsymbol{x},\boldsymbol{y}) \coloneqq a_r(\boldsymbol{y})(\mathbb{I}_{\boldsymbol{y}+\boldsymbol{v}_r}(\boldsymbol{x})-\mathbb{I}_{\boldsymbol{y}}(\boldsymbol{x}))$, where $\boldsymbol{v}_r := (v^{+}_{rs}-v^{-}_{rs})^\intercal_{s \in \mathcal{S}'}$ is the stoichiometric vector and $\boldsymbol{x},\boldsymbol{y} \in \boldsymbol{\mathcal{X}}$ with $\boldsymbol{\mathcal{X}} \subseteq \mathbb{N}^{|\mathcal{S}'|}$ denoting the full state space. In the following, we assume that the state space is irreducible; hence, there exists a unique solution to the CME \citep{doi:10.1137/19M1289625}.

\begin{figure}[t!]
\centering
\includegraphics[width=\columnwidth]{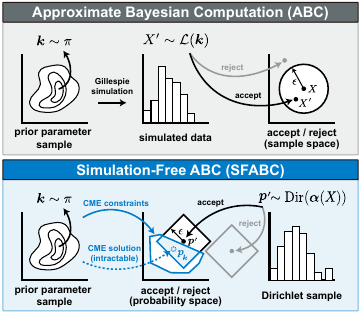}
\caption{\textbf{Comparison of ABC and SFABC for SRNs.} ABC simulates data $X'$ from the CME likelihood $\mathcal{L}$ using the Gillespie algorithm given a parameter set $\boldsymbol{k}$ drawn from the prior $\pi$. Parameters producing simulations $X'$ close ($\epsilon$-neighborhood) to the observed data $X$ are accepted to approximate the posterior distribution (\textit{SI Text 1}). SFABC draws probability distributions $\boldsymbol{p}'$ using Dirichlet resampling of the data. 
A parameter set $\boldsymbol{k}$ is accepted if the $ \epsilon$-neighborhood of the candidate probability $\boldsymbol{p}'$ contains probability vectors that satisfy the CME constraints, which can be checked by convex optimization. The CME constraints present a relaxation containing the intractable CME solution $\boldsymbol{p}_{\boldsymbol{k}}$ that can be systematically refined. 
%A parameter set $\boldsymbol{k}$ is accepted if the CME constraints (relaxation) containing the CME solution $\boldsymbol{p}_{\boldsymbol{k}}$ overlaps with an $\epsilon$-neighborhood of the candidate probability $\boldsymbol{p}'$ as determined by convex optimization. %While these illustrations assume a steady-state dataset, both ABC and SFABC can handle transient data with multiple observation times.
}\label{fig:Fig1}
\end{figure}

Calculating the explicit state probability $\boldsymbol{p}_{\boldsymbol{k},t}$, and hence the likelihood, requires solving the CME in \eqref{eqn:CME}, which is often mathematically demanding because the underlying state space is typically infinite, giving rise to an infinite system of coupled differential equations. While likelihood-based MCMC is applicable only to the limited class of SRNs with tractable likelihoods, simulation-based methods such as ABC avoid explicit likelihood evaluation by using stochastic simulations. However, repeated simulations can become computationally prohibitive. We therefore introduce a simulation-free ABC framework that uses CME-derived constraints to directly assess the compatibility of candidate parameters with the observed data (see Fig.~\ref{fig:Fig1}), requiring neither stochastic simulation nor explicit likelihood evaluation.

\subsection{SFABC Provides Likelihood- and Simulation-Free Posterior Estimation} 

The core idea of SFABC is to accept (or reject) a prior parameter sample $\boldsymbol{k}$ based on whether the resampled data is compatible (or incompatible) with the solution of the CME. A computational challenge with this approach is that the CME typically evolves over an unbounded state space larger than the one observed in the data. We circumvent this issue by formulating a tractable convex linear optimization problem that can be solved using standard solvers \cite{andersen2000mosek,mosek}. 
Specifically, we consider only a finite underdetermined set of equations in \eqref{eqn:CME} involving probabilities over a subset $\boldsymbol{\mathcal{M}}\subseteq \boldsymbol{\mathcal{X}}$ of the state space. %that contains the observed states $\boldsymbol{\mathcal{U}}_t$ for all $t\in\mathcal{T}$. 
To deal with the time-derivatives involved in transient problems, we integrate \eqref{eqn:CME} against a range of suitable time-dependent test functions which provides constraints on a finite time grid %$\mathcal{T}_{\text{opt}}\coloneqq\{t_{i,\text{opt}}\}_{i=1}^{T'}\supseteq\mathcal{T}$ 
involving the observation times (see \textit{SI Text 2}). The resulting relaxation approximation represents a set $\boldsymbol{\mathcal{P}}_{\boldsymbol{k}}^{\boldsymbol{\mathcal{M}}}$ of plausible CME solutions given a finite set of equations that also contains the true solution and can be tightened by increasing the number of equations through $\boldsymbol{\mathcal{M}}$.

Let $\mathcal{T}\coloneqq\{t_i\}_{i=1}^T$ denote the set of unique observation times in $\boldsymbol{\tau}$. 
For each $t$, let $B_t$ be the number of unique observed states at time $t$ with $1\leq B_t \leq N_t$, and we denote the unique states by $\boldsymbol{\mathcal{U}}_t\coloneqq \{\boldsymbol{u}_{b,t}\}_{b=1}^{B_t}$ with numbers of occurrence $n_{1,t},\ldots,n_{B_t,t}$ satisfying $\sum_{b=1}^{B_t}n_{b,t}=N_t$. The SFABC rejection sampling algorithm is then defined as follows:
\vspace{1mm}
\begin{description}
    \item \textbf{SFABC rejection sampling (SFABC-Rej)} 
    \begin{enumerate} 
        \item[\textit{S1}] Sample a parameter set $\boldsymbol{k}$ from the prior $\pi$.
        \item[\textit{S2}] Sample probability distributions $\boldsymbol{p}'_t\sim\mathrm{Dir}(\boldsymbol{\alpha}(X_t))$ with $\boldsymbol{\alpha}(X_t)\coloneqq (n_{1,t}+1,\ldots,n_{B_t,t}+1, 1)^\intercal\in\mathbb{N}_+^{B_t+1}$ for each observation time $t\in\mathcal{T}$.
        \item[\textit{S3}] 
        Test if there exists a CME relaxation $\{\boldsymbol{p}_t\}_{t\in\mathcal{T}}\in\boldsymbol{\mathcal{P}}_{\boldsymbol{k}}^{\boldsymbol{\mathcal{M}}}$ that satisfies $\max_{t\in\mathcal{T}} d_{\boldsymbol{\mathcal{U}}_t}(\boldsymbol{p}_{t},\boldsymbol{p}'_t)\leq\epsilon$. If such a solution exists, accept $\boldsymbol{k}$. Otherwise, go to \textit{S1}.
        \item[\textit{S4}] Weigh the accepted sample $\boldsymbol{k}$ by $1/V_{\epsilon}^{\boldsymbol{\mathcal{M}}}(\boldsymbol{k})$ where the volume of acceptance region $V_{\epsilon}^{\boldsymbol{\mathcal{M}}}(\boldsymbol{k})\coloneqq \mathbb{E}_{\{\bar{\boldsymbol{p}}_t\sim\text{Dir}(\boldsymbol{1}(X_t)\}_{t\in\mathcal{T}}}\left[\mathbb{I}_{\boldsymbol{\mathcal{P}}_{\boldsymbol{k}}^{\boldsymbol{\mathcal{M}}}}(\max_{t\in\mathcal{T}} d_{\boldsymbol{\mathcal{U}}_t}(\boldsymbol{p}_t,\bar{\boldsymbol{p}}_t)\leq\epsilon)\right]$, where $\boldsymbol{1}(X_t)\coloneqq(1,\ldots,1)^\intercal\in\mathbb{N}_+^{B_t+1}$.
    \end{enumerate}
\end{description} 
\vspace{1mm}
Here, \textit{S1} samples a parameter $\boldsymbol{k}$ from the prior.
In \textit{S2}, we sample a probability distribution from a Dirichlet distribution whose first $B_t$ components correspond to the probabilities over the observed states $\boldsymbol{\mathcal{U}}_t$, and the last component aggregates the total probability mass over all unobserved states. This step captures finite-sample variability. Step \textit{S3}, called the feasibility test, tests whether the probability distributions $\boldsymbol{p}'_t$ for all $t\in \mathcal{T}$ are close (within an $\epsilon$ neighborhood) to a CME relaxation with parameters $\boldsymbol{k}$. 
%\ZL{Here, $d_{\boldsymbol{\mathcal{U}}_t}(\boldsymbol{p}_t,\boldsymbol{p}_t')$ is any chosen distance measure computed over the observed states $\boldsymbol{\mathcal{U}}_t$. An example is the $1$-norm distance $d_{\boldsymbol{\mathcal{U}}_t}(\boldsymbol{p}_t,\boldsymbol{p}_t')=\sum_{u \in \boldsymbol{\mathcal{U}}_t} |p_t(u) - p_t'(u)|$.}
The distance $d_{\boldsymbol{\mathcal{U}}_t}(\boldsymbol{p}_t,\boldsymbol{p}_t')=\sum_{u \in \boldsymbol{\mathcal{U}}_t} |p_t(u) - p_t'(u)|$ is the $1$-norm computed over the observed states $\boldsymbol{\mathcal{U}}_t$ or a suitably chosen distance of summary statistics (\textit{SI Text 4}). 
%\ZL{This is achieved by finding feasible points in the set $\boldsymbol{\mathcal{P}}_{\boldsymbol{k}}$ with linear constraints from the CME and the regularizing conditions on the probabilities.} 
Finally, \textit{S4} weighs the accepted parameters by the inverse volume of the acceptance region obtained by repeated sampling from uniform Dirichlet probability distributions (see \MM). A generalization of this algorithm is Sequential Monte Carlo (SFABC-SMC) that can improve sampling efficiency for parameter inference and model selection (\textit{SI Text 3}).

The construction is a computationally tractable linear program relaxation of the CME solution that converges to it as the subset of equations covers the whole probability distribution and temporal dimension, namely as $\boldsymbol{\mathcal{M}}\rightarrow \boldsymbol{\mathcal{X}}$ and the time discretization decreases to zero. The accepted samples in SFABC-Rej therefore can, in principle, approximate the true posterior \eqref{eqn:posterior} arbitrarily closely as $\epsilon \to 0$ (see convergence proof in \MM ). In practice, the resulting posterior approximations are expected to be wider than the limiting true posteriors since the relaxation approximation contains the true CME solution. In the following, we validate that SFABC provides accurate posterior approximations and demonstrate how it can be adapted to a broad range of single-cell transcriptomic data. 
 
%In practice, however, the feasibility problem in \textit{S3} is generally infinite-dimensional with an infinite number of equations in $\boldsymbol{\mathcal{P}}_{\boldsymbol{k}}$ involving probabilities over an unbounded state space $\mathcal{X}$ and their time derivatives.
%the acceptance condition in \textit{S3} amounts to checking whether the candidate probability is close to a solution of the CME under the given parameter settings. However, this feasibility problem is generally infinite-dimensional with an infinite number of equations involving probabilities over an unbounded state space $\mathcal{X}$ and their time derivatives. 

\subsection{SFABC Accurately and Robustly Estimates the Posterior Distribution with Lower Computational Cost}
\begin{figure*}[t!]
    \centering
    \includegraphics{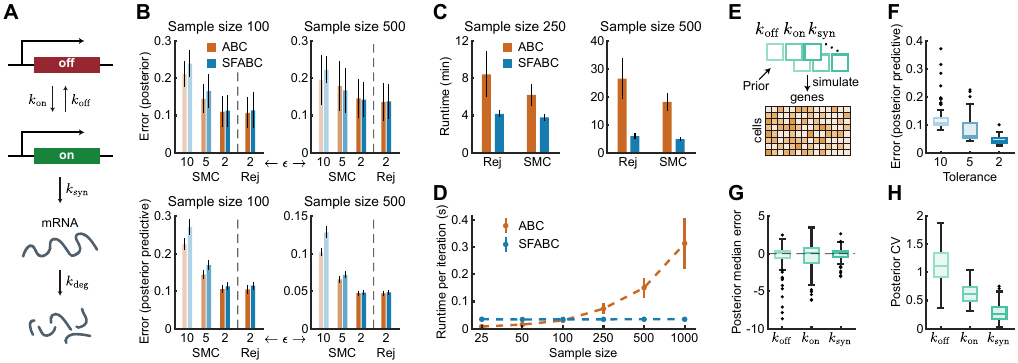}
    \caption{\textbf{Validation of SFABC with Steady-State Synthetic Data from the Telegraph Model.} \textbf{(A)} Illustration of the telegraph model. \textbf{(B)} Validation of SFABC algorithms with sample sizes 100 and 500 drawn from the analytical steady-state distribution \cite{raj2006stochastic} ($k_{\text{off}}=2$, $k_{\text{on}}=8$, $k_{\text{syn}}=5$, $k_{\text{deg}}=1$). For each sample size, 20 datasets are simulated, and 500 posterior samples are collected per dataset using MCMC, ABC-Rej, ABC-SMC, SFABC-Rej, and SFABC-SMC sampling algorithms. The SMC algorithms use tolerance sequence $\epsilon = \{10,5,2\}$ and the rejection samplers use $\epsilon=2$, and other inference setups are provided in S1 - S4 in \textit{SI Table~S1}. Top panels: mean Kolmogorov--Smirnov (KS) statistic between the marginal distributions of SFABC posterior samples and MCMC samples. Bottom panels: mean KS statistic between each observed dataset and 500 synthetic datasets drawn from the corresponding posterior predictive distribution. Bars indicate mean metric $\pm$ standard deviation. \textbf{(C)} Total wall time for sample sizes 250 and 500. Experiments are repeated over 20 datasets, and bars indicate the mean time $\pm$ standard deviation. \textbf{(D)} Wall time per iteration (simulation for ABC and feasibility test for SFABC) with sample sizes from 25 to 1,000. Points and bars indicate mean wall time $\pm$ standard deviation over 100 iterations. \textbf{(E)} Validation of SFABC inference across parameter sets. Parameters $k_{\text{off}}, k_{\text{on}}, k_{\text{syn}}$ are drawn independently from $10^{\text{Uniform}(-1,1)}$ and used to simulate datasets of size 1,000 from the steady-state distribution. We retain only datasets with mean expression greater than 0.5, yielding 100 parameter sets and synthetic datasets. For each dataset, we use SFABC-SMC to collect 100 posterior samples with tolerance sequence $\epsilon = \{10,5,2\}$ (full setup in S5 in \textit{SI Table~S1}). \textbf{(F)} KS statistic between observed data and 100 synthetic datasets drawn from the posterior predictive distribution. \textbf{(G)} Difference between the true parameter and posterior median with $\epsilon=2$. \textbf{(H)} Coefficient of variation (CV) of marginal posterior distributions with $\epsilon=2$.}
    \label{fig:Fig2}
\end{figure*}

First, we validate the accuracy of SFABC's posterior estimate against exact posterior samples obtained using a likelihood-based MCMC sampler (see \textit{SI Text 4}) and we benchmark its performance against standard simulation-based ABC. %
We consider synthetic snapshot data from the telegraph model (see \textit{SI Text 5} and illustration in Fig.~\ref{fig:Fig2}A). At steady state, the time derivative in \eqref{eqn:CME} is zero, which reduces the CME to algebraic constraints:
\begin{equation} \label{eqn:CME-ss}
    \sum_{r=1}^R k_r Q_r \boldsymbol{p}_{\boldsymbol{k}} = 0.
\end{equation} 
Restricting the above to a finite subset of equations involving the observed state probabilities and the distance to the sampled histogram then provides constraints that represent our relaxation approximation (\textit{SI Text 2}). 

%The telegraph model has $R=4$ rate parameters: the gene state switching rates $k_{\text{off}}$ and $k_{\text{on}}$, the mRNA production rate $k_{\text{syn}}$, and and the mRNA degradation rate $k_{\text{deg}}$. We fix $k_{\text{deg}}\equiv1$ due to lack of identifiability at steady state and infer the remaining three parameters with independent Gamma priors.

%Accuracy paragraph
We find that SFABC accurately estimates the posterior distribution across datasets with varying sample sizes (Fig.~\ref{fig:Fig2}B). The approximation errors of SFABC-SMC decrease with tolerance and achieve accuracy similar to that of SFABC-Rej, consistent with SFABC's guarantee of convergence to the true posterior (blue bars). This is quantified by the Kolmogorov–Smirnov (KS) distance between the exact posterior obtained by MCMC and the SFABC approximation (top), and similarly, the KS distance between the snapshot data distribution and SFABC's posterior predictive distribution (bottom row). Posterior accuracy improves with the number of equations used in the optimization and as the tolerance decreases, but this also decreases the acceptance probability as expected (\textbf{\textit{SI Fig.~S1}}). This accuracy is comparable to that of the standard ABC (Fig.~\ref{fig:Fig2}B, red bars), despite fundamental differences in the rejection step, and is independently verified for our SMC implementations, indicating that the relaxation employed by SFABC (\textit{SI Text 4}) provides an excellent approximation to the CME. 

%Computational cost paragraph
While having similar accuracy, SFABC-Rej achieves a marked 50\% reduction in computation time for sample size 250 and a 78\% reduction for sample size 500 compared to ABC-Rej, and a similar scaling advantage is observed with the corresponding SMC samplers (Fig.~\ref{fig:Fig2}C). The reduction in computational cost stems from the fact that the feasibility test with an LP solver at each SFABC iteration is virtually independent of sample size, in contrast to ABC's simulation step (Fig.~\ref{fig:Fig2}D). 

%When the truncated state space is much smaller than the data support, however, the feasibility constraints are too weak, and SFABC samples only from the prior.

%Robustness paragraph
Next, we assess the robustness of SFABC-SMC across a wide range of parameter settings.  Across synthetic datasets obtained from 100 parameter sets sampled independently from an uninformative prior distribution (Fig.~\ref{fig:Fig2}E, \textbf{\textit{SI Fig.~S2}}A), SFABC produces accurate posterior predictive distributions and unbiased parameter estimates, with improved agreement as the tolerance decreases (Fig.~\ref{fig:Fig2}F-G). We observe that the pairwise posterior distributions from SFABC closely match those from ABC across datasets (\textbf{\textit{SI Fig.~S2}}B-D). Despite identical priors, the relative posterior uncertainty is significantly lower for $k_{\text{syn}}$ than for $k_{\text{on}}$ and $k_{\text{off}}$ (Fig.~\ref{fig:Fig2}H). This is consistent with the intuition that $k_{\text{syn}}$ sets expression levels \cite{tang2023modelling}. Our results demonstrate the correctness, performance, and robust inference of SFABC.

\subsection{SFABC Enables Bayesian Analysis of Genome-wide Burst Kinetics}
Next, we apply SFABC to analyze two scRNA-seq datasets in mouse fibroblast cells that differ substantially in sample size (224 cells in D1 \cite{larsson2019genomic}; 8,916 in D2 \cite{ramskold2024single}; \textit{SI Text 6}), and infer the underlying burst kinetics using the telegraph model. SFABC-SMC is calibrated with the tolerance sequences (setups S7 and S8 in \textit{SI Table~S1}) selected based on posterior predictive checks on a subset of genes (\textbf{\textit{SI Fig.~S3}}).

\begin{figure*}[ht]
\centering
\includegraphics{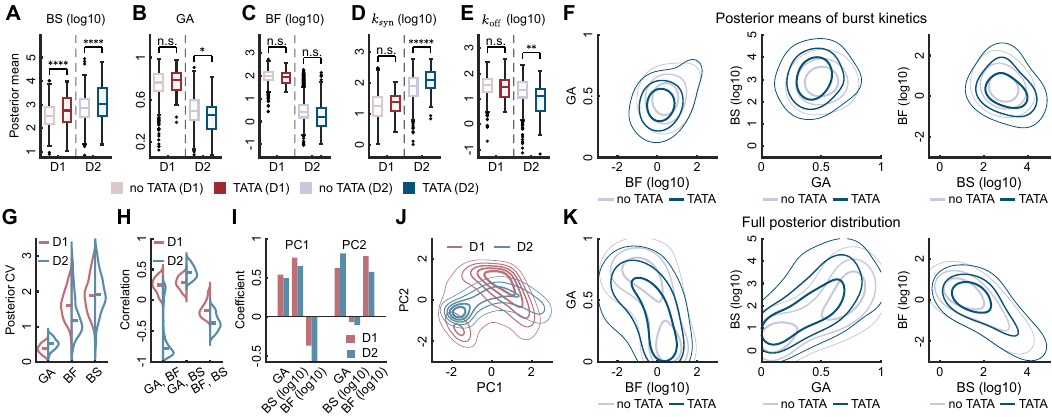}
\caption{\textbf{Inference of Burst Dynamics using the Telegraph Model from Snapshot scRNA-seq Data.} \textbf{(A)-(E)} Posterior means of burst kinetics and individual rate parameters of the telegraph model (see cartoon in Fig.~\ref{fig:Fig2}A), grouped by mouse fibroblast data (D1 \cite{larsson2019genomic}, D2 \cite{ramskold2024single}) and promoter type (with or without a TATA element). \textbf{(F)} Smoothed contour plots of posterior means of gene activity ($\text{GA}=k_{\text{on}}/(k_{\text{on}}+k_{\text{off}})$), burst size ($\text{BS}=k_{\text{syn}}/k_{\text{off}}$), and burst frequency ($\text{BF}=k_{\text{on}}/k_{\text{deg}}$) across genes (100 posterior samples per gene) in dataset D2, restricted to the 2.5th-97.5th percentiles. \textbf{(G)-(H)} CVs of marginal posterior distributions and pairwise correlations in posterior samples across genes in each dataset, restricted to the 10th-90th percentiles. \textbf{(I)} Coefficients of principle components. For each gene, principal component analysis is performed on the (across-gene) posterior samples of GA, $\log_{10}(\text{BS})$, and $\log_{10}(\text{BF})$ for each of D1 and D2. \textbf{(J)} Distribution of posterior samples projected to the plane of the first principal component (PC1) and the second one (PC2). \textbf{(K)} Smoothed contour plots of the full posterior distribution of GA, BF, and BS across genes in dataset D2, restricted to the 2.5th-97.5th percentiles. In all panels, SFABC-SMC is applied to 979 filtered genes (\textit{SI Text 6}).}
\label{fig:Fig3}
\end{figure*}

We first examine how the kinetic parameters depend on the promoter architecture. As reported previously \cite{larsson2019genomic}, genes with a TATA-binding element in their core promoter exhibit significantly higher burst sizes ($\text{BS}=k_{\text{syn}}/k_{\text{off}}$) than non-TATA-box genes, and are therefore considered noisier. We observe the same pattern in the SFABC posterior means for D1 ($P=1.73.01\times10^{-4}$) and D2 ($P=9.58\times10^{-11}$) (Fig.~\ref{fig:Fig3}A). While the burst frequency ($\text{BF}=k_{\text{on}}/k_{\text{deg}}$ with $k_{\text{deg}}\equiv1$) shows no difference in both datasets (Fig.~\ref{fig:Fig3}C), several differences emerge in D2 that are not statistically significant in D1 due to low number of cells. These include lower gene activity (defined as $\text{GA}=k_{\text{on}}/(k_{\text{on}}+k_{\text{off}})$) for TATA-box genes ($P=1.69\times10^{-2}$) (Fig.~\ref{fig:Fig3}B), and higher BS are reflected by difference between promoter types in D2 such a higher $k_\text{syn}$ ($P=6.29\times10^{-6}$) and lower $k_{\text{off}}$ ($P=8.60\times10^{-3}$) for TATA-box genes, while no difference in individual rates is observed in D1 (Fig.~\ref{fig:Fig3}D-E). These findings suggest an additional role for TATA-containing promoters in modulating gene activity via promoter switching and thereby shaping overall burst kinetics distributions (Fig.~\ref{fig:Fig3}F).

Although D2 contains many more cells than D1, the posterior uncertainty of the burst kinetics BS, GA, and BF does not decrease substantially (Fig.~\ref{fig:Fig3}G). These measures could partly reflect differences between allele-specific expression in D1 and gene-level, effectively biallelic expression in D2 (\textit{SI Text 6}), rather than posterior uncertainty alone. To investigate this, we first examine gene-level posterior dependence and find that D2 yields much stronger pairwise correlations (Fig.~\ref{fig:Fig3}H). To assess whether this reflects a broader global structure in parameter identifiability, we next analyze the genome-wide posterior distribution that samples parameters across all genes. Principal component analysis of the burst kinetics (coefficients in Fig.~\ref{fig:Fig3}I) shows that the first principal component explains more than 70\% of the variance in D2, compared with only about 50\% in D1 (Fig.~\ref{fig:Fig3}J). The same pattern of posterior uncertainty was observed when D2 was subsampled to match the cell number in D1 (\textbf{\textit{SI Fig.~S4}}), suggesting that the first principal component quantifies uncertainty due to parameter identifiability, whereas the second component quantifies uncertainty due to finite sample size. Thus, sequencing more cells does not reduce posterior uncertainty but instead confines the posterior more strongly to a lower-dimensional manifold of burst kinetics, a behavior characteristic of the sloppiness in parameter identifiability \citep{gutenkunst2007universally}. 

%, in which only parameter combinations are tightly constrained, while uncertainty remains along weakly identifiable directions . 

The sloppiness of burst kinetics is supported by comparing gene-level posterior means with the full posterior distributions that aggregate posterior samples across all genes. When considering posterior means (Fig.~\ref{fig:Fig3}F), GA shows positive associations with BF and BS in both TATA-box and non-TATA-box genes (Pearson correlation coefficients $r_{\text{GA},\log_{10}(\text{BF})}=0.64$ and $r_{\text{GA},\log_{10}(\text{BS})}=0.20$ for TATA-box genes, and $0.53$ and $0.15$ for non-TATA-box genes).  In contrast, BF and BS exhibit a weak inverse relationship ($r_{\log_{10}(\text{BF}),\log_{10}(\text{BS})}=-0.34$ for TATA-box genes and $-0.37$ for non-TATA-box genes), suggesting coordinated regulation of bursting kinetics at the between-gene level. In contrast, analysis of all posterior samples across genes reveals a stronger inverse relationship between BF and BS ($r_{\log_{10}(\text{BF}),\log_{10}(\text{BS})}=-0.79$ for TATA-box genes and $-0.82$ for non-TATA-box genes), as well as positive and negative associations of GA with BS ($r_{\text{GA},\log_{10}(\text{BS})}=0.48$ for TATA-box genes and $0.64$ for non-TATA-box genes) and with BF ($r_{\text{GA},\log_{10}(\text{BF})}=-0.10$ for TATA-box genes and $-0.35$ for non-TATA-box genes), respectively (Fig.~\ref{fig:Fig3}K). These strong within-gene correlations differ from, and in some cases oppose, the trends observed in the posterior means (Fig.~\ref{fig:Fig3}F), suggesting weak identifiability of burst kinetics.

\begin{figure*}[t]
    \centering
    \includegraphics{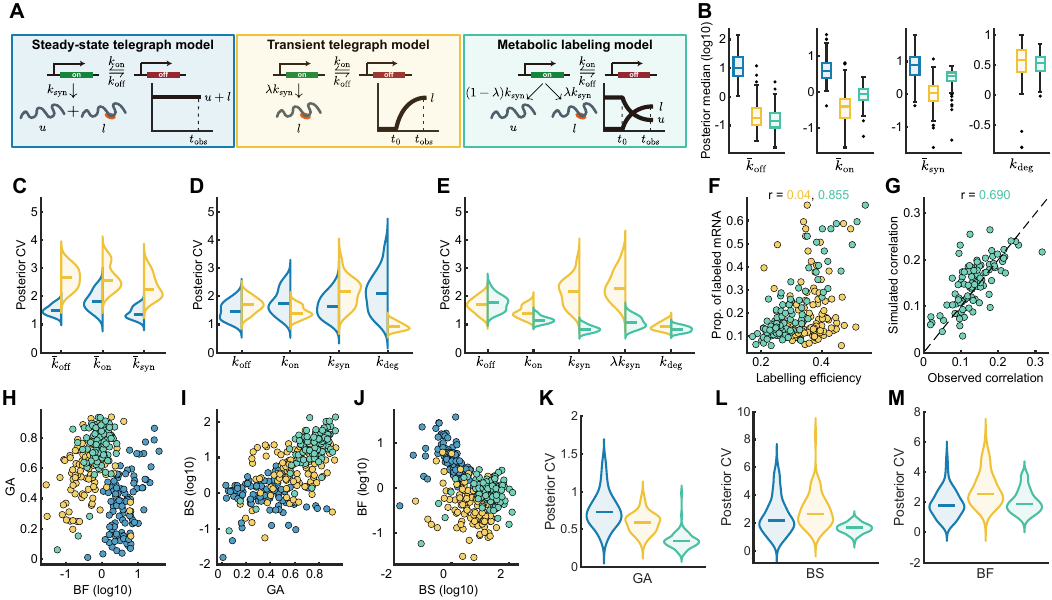}
    \caption{\textbf{Inference of Burst Kinetics and Labeling Efficiency Using Time-resolved Transcriptomics.} \textbf{(A)} Illustrations of stochastic reaction networks and data modalities. All models are fitted to primary mouse fibroblasts 4sU-labelled scRNA-seq data in Ref. \citep{ramskold2024single} that provides labeled and unlabeled counts $l$ and $u$ at time-point $t_{\text{obs}} = 2 h$. The steady-state telegraph model considers their sum $u+l$, the transient telegraph uses only the labeled count $l$, and the metabolic labeling model involves paired observations $(u,l)$. Degradation of labeled and unlabeled mRNA molecules occurs at a rate $k_{\text{deg}}$ in all models. 
    \textbf{(B)} Posterior median of the scaled switching rates $\bar{k}_{\text{on}}\coloneqq k_{\text{on}}/k_{\text{deg}}$ and $\bar{k}_{\text{off}}\coloneqq k_{\text{off}}/k_{\text{deg}}$, the scaled effective production rate $\bar{k}_{\text{syn}}\coloneqq k_{\text{syn}}/k_{\text{deg}}$ for steady-state telegraph and metabolic labeling models and $\bar{k}_{\text{syn}}\coloneqq \lambda k_{\text{syn}}/k_{\text{deg}}$ for transient telegraph model, and the degradation rate $k_{\text{deg}}$ in the transient models. In all three models, we use priors $k_{\text{off}},k_{\text{on}}\sim10^{\text{Uniform}(-1.5,1.5)}$ and $k_{\text{syn}}\sim 10^{\text{Uniform}(-1.5,3)}$. The prior of the labeling efficiency $\lambda$ is $10^{\text{Uniform}(-1,0)}$ in the transient models. Inference is performed on a subset of 100 genes (gene selection details in \textit{SI Text 6}), and inference setups are provided in S9 - S11 of \textit{SI Table~S1}. \textbf{(C)-(D)} Posterior CVs of scaled parameters and original parameters with steady-state and transient telegraph models. \textbf{(E)} Posterior CVs of original parameters with transient telegraph and metabolic labeling models. \textbf{(F)} Proportion of labeled mRNA in observed data, averaged over cells, versus the posterior mean of inferred labeling efficiency $\lambda$ in transient telegraph and metabolic labeling models. The Pearson correlation coefficient ($r$) between them is reported for each model. \textbf{(G)} Correlation between labeled and unlabeled counts in posterior predictive data against that in observed data. \textbf{(H)-(J)} Pairwise plots of posterior median of burst kinetics GA, BF, and BS (calculated from the scaled parameters $\bar{k}_{\text{on}}, \bar{k}_{\text{off}}, \bar{k}_{\text{syn}}$). \textbf{(K)-(L)} Posterior CVs of burst kinetics GA, BF, and BS.}
    \label{fig:Fig4}
\end{figure*}
%This discrepancy highlights that, while point estimates capture between-gene patterns, the full Bayesian posterior reveals strong within-gene parameter trade-offs, providing further evidence of weak identifiability in burst kinetics.

Notably, the full posterior reveals differences in gene activity for non-TATA-box and TATA-box genes. Non-TATA-box genes reveal a bimodal GA distribution (Fig.~\ref{fig:Fig3}K), suggesting that these genes occupy two distinct activity regimes, a feature that is obscured when considering only posterior means and also not observed for TATA-containing genes, which generally have lower GA (compare Fig.~\ref{fig:Fig3}B). This suggests that promoter architecture may affect not only bursting behavior but also the range of accessible transcriptional states, with non-TATA-box promoters exhibiting greater heterogeneity in gene activity than TATA-containing promoters. Together, these results highlight the importance of gene activity in modulating burst kinetics.

\subsection{SFABC Quantifies Absolute Burst Kinetics using Time-resolved Transcriptomics}

% Dataset D2 from \cite{ramskold2024single} uses metabolic labeling and contains both labeled ($l$) and unlabeled ($u$) mRNA counts. In the previous section, only the total mRNA counts were used to infer kinetic parameters at a relative time scale under a steady-state telegraph model with $k_{\text{deg}}\equiv 1$. 
Next, we use SFABC for uncertainty quantification in the time-resolved analysis of metabolic labeling data \cite{ramskold2024single}. We consider three models associated with different data modalities. The first is a steady-state telegraph model that uses only the total mRNA count ($u+l$ modality), the second is a transient telegraph model that describes the dynamics of newly labeled mRNA alone ($l$), and the third is a metabolic labeling model that decomposes transcription into labeled and unlabeled mRNAs ($u$ and $l$). See Fig.~\ref{fig:Fig4}A and \textit{SI Text 5}. We infer all four rate parameters $k_{\text{on}}$, $k_{\text{off}}$, $k_{\text{syn}}$, $k_{\text{deg}}$ and the labeling efficiency $\lambda$ under the same priors (see caption of Fig.~\ref{fig:Fig4}B). 
%. , assuming no labeled transcripts at $t=0$ and that they are produced with rate $\lambda k_{\text{syn}}$ where $\lambda$ is the labeling efficiency.  with production rates $\lambda k_{\text{syn}}$ and $(1-\lambda) k_{\text{syn}}$, respectively. In this model, no labeled mRNA is present at $t=0$, while the total mRNA population follows a steady-state distribution at all times.

% In this section, we investigate how transient metabolic labeling information enables inference of burst kinetics at an absolute time scale. We consider three models and their corresponding data modalities with labeled ($l$) and unlabeled ($u$) mRNA counts (Fig.~\ref{fig:Fig4}A). The first is a steady-state telegraph model that uses only the total mRNA count $u+l$, and we infer all four rate parameters. The second is a transient telegraph model that describes the dynamics of newly labeled mRNA alone, assuming no labeled transcripts at $t_0=0\text{h}$ and an observation of $l$ at $t_{\text{obs}}=2\text{h}$. The third is a metabolic labeling model (SRN in \textit{Materials and Methods}) that decomposes transcription into labeled and unlabeled mRNAs with production rates $\lambda k_{\text{syn}}$ and $(1-\lambda) k_{\text{syn}}$, respectively. In this model, no labeled mRNA is present at $t_0=0\text{h}$, while the total mRNA population follows a steady-state distribution at all times, and both labeled and unlabeled counts $(u,l)$ are observed at $t_{\text{obs}}=2\text{h}$.

We adapt SFABC to the time-resolved setting by transforming the transient CME \eqref{eqn:CME} into algebraic constraints that link the rate parameters with snapshot observations. To this end, we introduce an exponential time-weighting and integrate between time points $t_1$ and $t_2$ with $t_2> t_1\geq 0$ as in \citep{dowdy2018bounds}. The result is:
\begin{align} \label{eqn:CME-tran}
    e^{-\rho t_2}\boldsymbol{p}_{\boldsymbol{k},t_2} - e^{-\rho t_1}\boldsymbol{p}_{\boldsymbol{k},t_1} + \rho\boldsymbol{q}_{\boldsymbol{k},t_1,t_2;\rho}= \sum_{r=1}^R k_rQ_r \boldsymbol{q}_{\boldsymbol{k},t_1,t_2;\rho},
\end{align} 
where $\boldsymbol{q}_{\boldsymbol{k},t_1,t_2;\rho} \coloneqq \int_{t_1}^{t_2} e^{-\rho t}\boldsymbol{p}_{\boldsymbol{k},t} \mathrm{d}t$ is not observable as it involves missing information between observation times. The transient CME assumes initial conditions with zero labeled transcripts ($l$) and a steady-state number of unlabeled transcripts ($u$). The latter obeys the steady-state CME \eqref{eqn:CME-ss}, as does the total mRNA ($u+l$) at all times. The resulting relaxation can be efficiently solved using linear programming (see details in \textit{SI Text 2}).

%In contrast to the steady-state case with the exact dynamical constraint \eqref{eqn:CME-ss}, the time-resolved setting admits relaxed constraints due to missing transient information between observation times. To link the rate parameters with snapshot observations, we derive the following constraint:
%\begin{align} \label{eqn:CME-tran}
%    e^{-\rho t_2}\boldsymbol{p}_{\boldsymbol{k},t_2} - e^{-\rho t_1}\boldsymbol{p}_{\boldsymbol{k},t_1} + \rho\boldsymbol{q}_{\boldsymbol{k},t_1,t_2;\rho}= \sum_{r=1}^R k_rQ_r \boldsymbol{q}_{\boldsymbol{k},t_1,t_2;\rho},
%\end{align} 
%which is obtained by introducing an exponential weighting over time and integrating the CME between time points $t_1$ and $t_2$ with $t_2> t_1\geq 0$ \citep{dowdy2018bounds}, where $\boldsymbol{q}_{\boldsymbol{k},t_1,t_2;\rho} \coloneqq \int_{t_1}^{t_2} e^{-\rho t}\boldsymbol{p}_{\boldsymbol{k},t} \mathrm{d}t$ is not observable. 
%While this transient CME constraint is applied to the transient telegraph and metabolic labeling models, the steady-state constraint in \eqref{eqn:CME-ss} is used to set the initial condition of the total mRNA in the metabolic labeling model, and the resulting feasibility tests involve linear programming (see details in \textit{SI Text 3}). In the dataset considered in this section from \citep{ramskold2024single}, we have one observation at the time $t_{\text{obs}} = 2 (\text{h})$, and one can choose any set of time points in $[0,2]$.

% B
We used SFABC-SMC to infer the corresponding rate parameters of the three models and found that all models fit their respective data modalities (posterior predictive checks in \textbf{\textit{SI Fig.~S5A}}). We found that the inference of posterior medians for relative parameters varied substantially across models and data modalities (Fig.~\ref{fig:Fig4}B). Notably, the estimates from the metabolic labeling model tended to lie between those of the other two models, whereas degradation rates were consistent with those inferred from the transient telegraph model.

%Because parameter identifiability differs across models, we compare the identifiable parameters and find that their posterior medians vary systematically by model (Fig.~\ref{fig:Fig4}B). Notably, by integrating steady-state information from total counts with transient information from newly labeled counts, the metabolic labeling model produces posterior medians for the scaled parameters that tend to fall between those of the other two models, while yielding a posterior median for the degradation rate close to that of the transient telegraph model.

% We first compare the scaled parameters due to the lack of identifiability in the steady-state model. We define the scaled switching rates $\bar{k}_{\text{on}}\coloneqq k_{\text{on}}/k_{\text{deg}}$ and $\bar{k}_{\text{off}}\coloneqq k_{\text{off}}/k_{\text{deg}}$ in all models, and 
% scaled effective production rate $\bar{k}_{\text{syn}}\coloneqq k_{\text{syn}}/k_{\text{deg}}$ in the steady-state telegraph and the metabolic labeling models and $\bar{k}_{\text{syn}}\coloneqq \lambda k_{\text{syn}}/k_{\text{deg}}$ for newly labeled mRNAs in the transient telegraph model. By combining steady-state information from total counts with transient information from newly labeled counts, the metabolic labeling model yields posterior medians of the scaled parameters that tend to interpolate between those of the other two models, while identifying the degradation rate with a similar posterior mean to that of the transient telegraph model (Fig.~\ref{fig:Fig4}B). 

We found that parameter identifiability also varied substantially across models. Posterior uncertainty of the transient telegraph model was significantly worse than in the steady-state case (Fig.~\ref{fig:Fig4}C). The telegraph model clearly identified degradation rates, but the identifiability of the other absolute rate parameters did not significantly improve over the steady-state telegraph model (Fig.~\ref{fig:Fig4}D), even though the latter can only partially identify these absolute parameters due to its snapshot nature. The metabolic labeling model, which integrates labeled and unlabeled counts, yielded tighter posteriors for all parameters on an absolute timescale except $k_\text{on}$, and thus provided more reliable inference (Fig.~\ref{fig:Fig4}E). Improved identifiability in the metabolic model was due to incorporating labeling efficiency $\lambda$, which was predictive of the labeled mRNA fraction $\lambda$ (Fig.~\ref{fig:Fig4}F). Metabolic labeling model simulations also recovered the labeled-unlabeled correlations across genes ($r = 0.69$, Fig.~\ref{fig:Fig4}G), validating that the model successfully integrates both transient and snapshot data modalities. 

Data modality had a marked effect on inferred burst kinetics (Fig.~\ref{fig:Fig4}H–J). Although burst size and burst frequency exhibited the expected inverse relationship, the steady-state and transient telegraph models inferred a broad range of gene activities and burst sizes, with median burst parameters strongly correlated across genes. By contrast, the metabolic-labeling model inferred uniformly higher gene activity with substantially reduced gene-to-gene variability. The identifiability gain was accompanied by positive correlations between degradation rate and both gene activity and burst size, which were less pronounced in the transient model (\textbf{\textit{SI Fig.~S5D–F}}). Consistent with these observations, the metabolic-labeling model substantially reduced posterior uncertainty in gene activity and burst size relative to the other models and data modalities (Fig.~\ref{fig:Fig4}K–M). These results highlight that exploiting the joint distribution of time-resolved data significantly improves the identifiability of transcriptional burst kinetics.

% F
%We observe significant shifts in inferred burst kinetics when transient information is incorporated (Fig.~\ref{fig:Fig4}H-J). Relative to the steady-state model, the transient telegraph model yields higher gene activity and burst size but lower burst frequency. These effects are amplified in the metabolic labeling model, accompanied by a general reduction in variability in burst kinetics across genes. \ZL{In addition to shifted posterior medians of burst kinetics, the metabolic labeling model produces posterior estimates with improved uncertainty (\ref{fig:Fig4}K-M).} Examining relationships with the degradation rate further clarifies these shifts (\textbf{\textit{SI Fig.~S4D-F}}). In the transient telegraph model, gene activity and burst size show moderate positive correlations with $k_{\text{deg}}$ ($r_{\text{GA},\log_{10}(k_{\text{deg}})}=0.20$ and $r_{\log_{10}(\text{BS}),\log_{10}(k_{\text{deg}})}=0.40$), whereas burst frequency exhibits a stronger negative correlation ($r_{\log_{10}(\text{BF}),\log_{10}(k_{\text{deg}})}=-0.62$). In the metabolic labeling model, the positive correlations of $k_{\text{deg}}$ with gene activity and burst size become more apparent ($r_{\text{GA},\log_{10}(k_{\text{deg}})}=0.36$ and $r_{\log_{10}(\text{BS}),\log_{10}(k_{\text{deg}})}=0.45$), while the negative correlation with burst frequency remains but is less pronounced ($r_{\log_{10}(\text{BF}),\log_{10}(k_{\text{deg}})}=-0.19$). Overall, leveraging full data modalities shifts the inferred burst kinetics and reveals clearer relationships between them and the degradation rate.

\subsection{SFABC Identifies Cell Cycle-Dependent Transcriptional Regulation}

\begin{figure*}[t]
\centering
\includegraphics{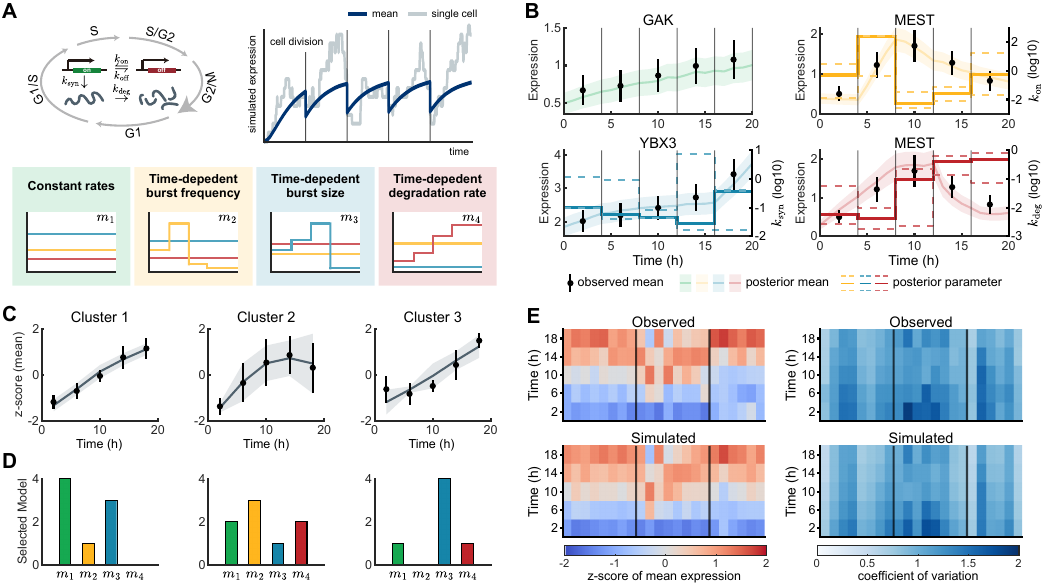}
\caption{\textbf{Model Selection among Telegraph Models with Cell Cycle-Dependent Rates.} \textbf{(A)} Cartoon of cell cycle-dependent telegraph models. Model simulations indicate that expression approximately doubles over cell cycles on average, with significant fluctuations from cell to cell and over time. The model with constant rates $m_1$ is the standard telegraph model, while $m_{2}$, $m_{3}$, and $m_{4}$ are telegraph models with piecewise constant burst frequency ($k_{\text{on}}$ varying), burst size ($k_{\text{syn}}$ varying), and degradation rate ($k_{\text{deg}}$ varying), respectively. \textbf{(B)} Representative gene-level fits and inferred time-dependent rate parameters to scRNA-seq data of RPE-1 cells with FUCCI cell cycle reporter (Ref. \cite{battich2020sequencing}; \textit{SI Text 6}). We use assume  parameters change discretely at timepoints $\{4,8,12,16\}$, experimental observation taken at $\mathcal{T}=\{2,6,10,14,18\}$ and cell division occurs at $t_\text{div}=20$ hours. For each example gene, black points show the observed mean mRNA counts at the sampled cell-cycle times, with error bars denoting the mean $\pm\ 0.2\times$ standard deviation of the observed counts. Coloured solid curves show the posterior predictive mean expression from stochastic simulations of the selected telegraph model, and shaded bands show $\pm\ 0.2\times$ the posterior predictive standard deviation. For models with a time-dependent parameter, the right y-axis shows the inferred $\log_{10}$ rate as a piecewise-constant function over the cell-cycle intervals; solid and dashed step curves denote the posterior median and the 95\% credible interval, respectively. \textbf{(C)} Z-scores of the mean expression level in each cluster. Black points and error bars show the mean $\pm\ 2$ standard deviations across the observed gene profiles, while the grey curves and shaded regions show the mean $\pm\ 2$ standard deviations of the z-score obtained from the simulated posterior predictive. \textbf{(D)} Model selection summary for fitted genes in each cluster. For each gene, posterior model probabilities are computed from the accepted SFABC samples, and the selected model is defined as the model with the largest posterior probability (\textit{SI Text 3}). Bars show the number of fitted genes assigned to each of the four candidate telegraph models within each cluster. \textbf{(E)} Observed and simulated z-scores of the mean expression level (left) and values of the coefficient of variation (right) per gene at the observation times. Vertical lines separate clusters. In (B) and (E), the simulation is run for 20 cell cycles to reach steady behavior before collecting posterior predictive samples for each parameter set drawn from the posterior. For each gene, 5,000 realizations from the posterior predictive distribution are simulated to plot the fitted expressions or to calculate the statistics. In all panels, we apply SFABC-SMC for model selection (\textit{SI Text 3}).}
    \label{fig:Fig5}
\end{figure*}

%An important question is how to distinguish between competing mechanistic hypotheses for transcriptional regulation. In particular, 
Transcriptional kinetics often vary across the cell cycle. As these effects are often ignored in transcriptome-wide studies, such undulations can be mistakenly reported as biological noise. Identifying the transcriptional processes that are regulated by the cell cycle is crucial for this purpose \citep{skinner2016single, liu2017transcriptional, liu2023slam}. To address this, we analyze scRNA-seq data with cell cycle reporters in \cite{battich2020sequencing} and use SFABC to perform model selection among telegraph models with explicit cell division and cell cycle-dependent rates. We consider four cell cycle-dependent models: constant rates and models with varying $k_{\text{on}}$, $k_{\text{syn}}$ or $k_{\text{deg}}$ (illustrations in Fig.~\ref{fig:Fig5}A).

To model the cell cycle progression, we use the transient equations \eqref{eqn:CME-tran} and assume parameters change discretely between observation times, the latter are akin to the phases (G1, G1/S, S, S/G2, G2/M) and cell division (M) at $t_\text{div}$. We then impose  
\begin{equation}
\boldsymbol{p}_{\boldsymbol{k},0} = B \boldsymbol{p}_{\boldsymbol{k},t_{\text{div}}},
\end{equation}
modeling equal partitioning of molecules at cell division. The matrix $B_{x,x'}=\frac{1}{2^{x'}}{ x' \choose x}$ represents the probability for a cell to inherit $x$ transcripts of the $x'$ transcripts of the mother cell \cite{rosenfeld2005gene,huh2011random}. 

%We clustered transcripts using $k$-means using z-scores of expression levels across the cell cycle. 
We applied SFABC-SMC for model selection to a number of representative genes with distinct temporal profiles (Figure \ref{fig:Fig5}B). To this end, we identified patterns by $k$-means clustering of the data (\textit{SI Text 6}), showing steadily increasing, plateauing, and delayed transcription profiles (Fig.~\ref{fig:Fig5}C, dots). SFABC recovered the mean expression levels for each cluster (lines), and its credible intervals recapitulated the gene distribution (Fig.~\ref{fig:Fig5}C, shaded). Model selection revealed that the clusters correspond to different rate modulations, with the constant models dominating the steadily increasing transcripts (cluster 1), while plateaued and delayed expression were associated with cell-cycle-dependent burst frequency (cluster 2) and burst size modulation (cluster 3, Fig.~\ref{fig:Fig5}D), respectively. The latter profiles (cluster 2 \& 3) were also associated with constant rates or cell-cycle-dependent degradation in some genes, indicating substantial heterogeneity in burst kinetics modulation (\textbf{\textit{SI Fig.~S6}}). 

%The mean expression of the fitted genes in each cluster and that from the posterior predictive is shown in Fig.~\ref{fig:Fig5}C. For clusters 1 and 2, the posterior predictive reproduces the temporal trends in the observed mean expression, while the fit is weaker for cluster 3 but still generally captures the overall pattern. The selected models for the fitted genes in each cluster are shown in Fig.~\ref{fig:Fig5}D. In cluster 1, the constant-rate model $m_1$ becomes the most selected one, indicating that the expression dynamics of the corresponding genes can be explained without explicit cell cycle-dependent modulation of transcriptional rates. By contrast, genes in clusters 2 and 3 favor models with time-dependent rates, with different rate modalities contributing across clusters. Moreover, within these clusters, different genes are best explained by different cell cycle-dependent models, indicating substantial heterogeneity in the underlying kinetic modulation (\textbf{\textit{SI Fig.~S5}}).

Using the best models, SFABC's predictive posterior distributions accurately recovered distinct cell-cycle patterns, and credible intervals recapitulated the data uncertainty (Figure \ref{fig:Fig5}B). In some cases (MEST), transcript profiles were equally well fit by different models with different rate modulations, indicating identifiability issues between cell cycle-dependent burst frequency and degradation. Overall, the posterior simulations reproduce the cluster-specific cell cycle patterns across genes (Fig.~\ref{fig:Fig5}E). The agreement indicates that the inferred models capture key temporal patterns in both mean expression and coefficient of variation across genes. Together, these results demonstrate that extending SFABC accurately captures biological noise and enables model comparison to unravel cell-cycle-dependent burst kinetics and transcriptional regulation.

\section{Discussion}
In this work, we propose SFABC, a simulation-free approximate Bayesian computation framework for stochastic reaction networks. SFABC leverages the Chemical Master Equation (CME), which links reaction-rate parameters to the underlying probability distributions of the stochastic reaction network. Rather than relying on stochastic simulations or analytical solutions, SFABC introduces a Bayesian sampling scheme that tests whether the CME associated with a given prior-parameter sample is compatible with Dirichlet-resampled data. These compatibility tests are formulated as convex optimization problems and can therefore be solved efficiently using linear programming.

We prove that the approximate posterior obtained by SFABC converges to the true Bayesian posterior in the limit of vanishing tolerance and increasingly refined relaxation approximations (see \MM). We further show that SFABC provides accurate posterior estimates from both snapshot and time-resolved transcriptomic data. Across synthetic benchmarks, SFABC achieves inference accuracy comparable to established ABC approaches while substantially reducing computational cost at large sample sizes, and remains robust across a wide range of parameter settings. When applied to steady-state scRNA-seq data, SFABC identifies systematic differences in inferred rate parameters associated with TATA-containing promoters and reveals pronounced within-gene parameter correlations, highlighting sloppiness and non-identifiability in burst kinetics. For time-resolved metabolic-labeling data, we find that identifiability is significantly improved by inferring absolute rate parameters and by exploiting correlation patterns across data modalities. Finally, we demonstrate the use of SFABC for model selection by identifying cell cycle-dependent transcriptional regulation.

In these applications, SFABC is able to incorporate model assumptions in a fundamentally different way from simulation-based approaches such as standard ABC. In the metabolic-labeling model, the total mRNA count is assumed to be at steady state. Similarly, the cell cycle-dependent telegraph model assumes cyclo-stationary behavior across cell divisions. Enforcing such stationary assumptions in simulation-based inference is computationally costly, since stochastic simulations must be run for sufficiently long times, or across sufficiently many cell cycles, to approach the required stationary behavior. Alternatively, dedicated simulators must be employed \citep{hemberg2008dominated}. In contrast, within our simulation-free approach, these assumptions translate into directly checkable equations, thereby reducing the computational burden associated with biological constraints.

SFABC also differs conceptually from rejection-based ABC in its use of distance functions and summary statistics. In SFABC, the distance function and summary statistics are computed over observed states, which act as test functions distinguishing candidate probability distributions. As a result, questions of statistical sufficiency arise in a technically different way from simulation-based frameworks, where distances are typically computed between simulated and observed summary statistics. However, issues of model mispecification and model selection may carry over to SFABC \cite{ratmann2009model,robert2011lack}.

Another fundamental distinction from approximate-likelihood approaches lies in how approximations are introduced. Finite-state projection, moment approximations, and synthetic likelihoods can induce systematic parameter bias \citep{frohlich2016inference}. By contrast, the relaxation strategy employed by SFABC is constructed to contain the true parameters and probability solutions at all sample sizes, but does so at the cost of wider posteriors and therefore greater uncertainty. Hybrid approaches that combine SFABC with finite-state projection schemes could improve computational performance, but may introduce bias; the convergence properties of such methods would therefore need to be reassessed \citep{kuntz2019exit}.

Although SFABC provides an efficient and flexible framework with provable convergence guarantees, its computational efficiency has several limitations. Whereas the cost of simulation-based methods is primarily driven by repeated simulations and typically scales with factors such as sample size, the cost of SFABC is governed by the complexity of the underlying optimization problem. This complexity grows exponentially with the number of molecular species, limiting the scalability of SFABC to more complex networks. In addition, the convergence guarantees require enlargement of the state space in both steady-state and transient settings, as well as refinement of the time mesh in the transient setting. These refinements increase the number of optimization variables and introduce a practical trade-off between accuracy and computational cost, alongside the usual tolerance-dependent trade-off. Adaptive state-space exploration, which remove low-probability states, could help alleviate this computational burden \citep{dendukuri2025adaptive}.

These limitations suggest several directions for extending and improving SFABC. One natural extension is to replace the full CME with moment equations, so that the optimization problem scales with the number and order of moments rather than with the size of the state space. This could substantially improve scalability to higher-dimensional networks, albeit at the cost of sacrificing the convergence guarantees associated with Dirichlet sampling of the likelihood in the probability-based formulation (see \textit{Materials and Methods}). Another promising direction is to reinterpret the feasibility test through the lens of control theory, with the aim of characterizing feasible parameter regions more directly and reducing the need for CME relaxation approximations \citep{dowdy2018dynamic, sakurai2018bounding, holtorf2024tighter}. It would also be valuable to translate algorithmic developments from ABC into the SFABC framework, including adaptive sequential Monte Carlo algorithms \citep{del2012adaptive} and regression adjustments for posterior correction \citep{beaumont2002approximate}.

More broadly, SFABC reframes Bayesian inference for stochastic reaction networks by showing that repeated simulation is not always necessary as the governing equations themselves impose informative constraints on the data-generating distribution. By replacing simulated trajectories with optimization-based compatibility tests, SFABC turns model equations into tractable inferential objects and provides a principled way to quantify uncertainty without solving the CME or approximating the likelihood. This perspective opens a new route towards simulation-free inference in complex biological systems, where rich single-cell measurements are increasingly available but direct likelihood evaluation and large-scale simulation remain major bottlenecks. In this sense, SFABC presents a step toward a more general class of inference methods that exploit the structure of dynamical models directly. We expect that this paradigm will be useful for the rational design and reverse engineering of biological systems.

\section{Materials and Methods}

\subsection{Practical Computation of the Acceptance Region Volume}
SFABC weighs accepted samples by the inverse volume of the corresponding acceptance region. Due to independence between time points, the volume is  $V_{\epsilon}^{\boldsymbol{\mathcal{M}}}(\boldsymbol{k}) = \prod_{t\in\mathcal{T}} V^{\boldsymbol{\mathcal{M}}}_{\epsilon,t}(\boldsymbol{k})$ where $V^{\boldsymbol{\mathcal{M}}}_{\epsilon,t}(\boldsymbol{k}) \coloneqq \mathbb{E}_{\bar{\boldsymbol{p}}_t\sim\text{Dir}(\boldsymbol{1}(X_t))}\left[\mathbb{I}_{\boldsymbol{\mathcal{P}}_{\boldsymbol{k}}^{\boldsymbol{\mathcal{M}}}}( d_{\boldsymbol{\mathcal{U}}_t}(\boldsymbol{p}_{t},\bar{\boldsymbol{p}}_t)\leq \epsilon)\right]$ for $t\in\mathcal{T}$. %Here, the indicator function test if a feasible solution in the CME relaxation set $\boldsymbol{\mathcal{P}}_{\boldsymbol{k}}$ satisfies the distance condition $\max_{t\in\mathcal{T}} d(\boldsymbol{p}_{t},\boldsymbol{p}'_t;\boldsymbol{\mathcal{U}}_t)\leq\epsilon$, determined by convex optimization.}
To estimate $V^{\boldsymbol{\mathcal{M}}}_{\epsilon,t}(\boldsymbol{k})$ without repeated optimisation, we use a feasible solution of the CME relaxation $\boldsymbol{p}_{t} \in \boldsymbol{\mathcal{P}}_{\boldsymbol{k}}^{\boldsymbol{\mathcal{M}}}$, repeatedly sample $\bar{\boldsymbol{p}}_t$ from the uniform Dirichlet distribution $\text{Dir}(\boldsymbol{1}(X_t))$, and take the proportion of samples with $\mathbb{I}_{\boldsymbol{\mathcal{P}}_{\boldsymbol{k}}^{\boldsymbol{\mathcal{M}}}}(d_{\boldsymbol{\mathcal{U}}_t}(\boldsymbol{p}_{t},\bar{\boldsymbol{p}}_t)\leq\epsilon)=1$ as our estimate, i.e.,
$V^{\boldsymbol{\mathcal{M}}}_{\epsilon,t}(\boldsymbol{k}) \approx \mathbb{E}_{\bar{\boldsymbol{p}}_t\sim\text{Dir}(\boldsymbol{1}(X_t))}\left[\mathbb{I}_{\boldsymbol{\mathcal{P}}_{\boldsymbol{k}}^{\boldsymbol{\mathcal{M}}}}(d_{\boldsymbol{\mathcal{U}}_t}(\boldsymbol{p}_{t},\bar{\boldsymbol{p}}_t)\leq\epsilon)\right]$, which does not affect convergence since $\boldsymbol{p}_{t}\to \boldsymbol{p}_{\boldsymbol{k},t}$ as $\boldsymbol{\mathcal{M}} \to \boldsymbol{\mathcal{X}}$.

\subsection{Convergence Proof of SFABC-Rej}
The distribution of the accepted samples in SFABC-Rej is
\begin{align*}
&P_\text{SFABC}(\boldsymbol{k}| \{X_{t}\}_{t\in\mathcal{T}}) 
\nonumber\\& \propto \int_{\boldsymbol{\mathcal{P}}} \mathrm{d}\{\boldsymbol{p}'_t\}_{t\in\mathcal{T}} \frac{\mathbb{I}_{\boldsymbol{\mathcal{P}}_{\boldsymbol{k}}^{\boldsymbol{\mathcal{M}}}}(\max_{t\in\mathcal{T}} d_{\boldsymbol{\mathcal{U}}_t}(\boldsymbol{p}_{t},\boldsymbol{p}'_t)\leq \epsilon)}{V_{\epsilon}^{\boldsymbol{\mathcal{M}}}(\boldsymbol{k})} \times \nonumber\\ & \quad \quad \quad \ \prod_{t\in\mathcal{T}}\prod_{b=1}^{B_t}  p'_t(\boldsymbol{u}_{b,t})^{n_{b,t}}\ \pi(\boldsymbol{k}),
\end{align*}
where $\boldsymbol{\mathcal{P}}\coloneqq\boldsymbol{\mathcal{P}}_1\times\cdots\times\boldsymbol{\mathcal{P}}_T$ is the joint support of the Dirichlet distributions over observation times such that
$\boldsymbol{\mathcal{P}}_t\coloneqq\{ \boldsymbol{p}\in[0,1]^{B_t+1}\ | \ \sum \boldsymbol{p} = 1 \}$ for $t\in\mathcal{T}$. To show convergence to the posterior distribution, we assume that the CME \eqref{eqn:CME} admits a unique solution for every parameter vector $\boldsymbol{k}$ and we take the limits: (1) as $\boldsymbol{\mathcal{M}}\rightarrow\boldsymbol{\mathcal{X}}$ (and $\max_i(t_{i+1,\text{opt}}-t_{i,\text{opt}})\rightarrow0$ for transient data) the set ${\boldsymbol{\mathcal{P}}_{\boldsymbol{k}}^{\boldsymbol{\mathcal{M}}}}\to{\boldsymbol{\mathcal{P}}_{\boldsymbol{k}}^{\boldsymbol{\mathcal{X}}}}$ is a singleton $\{\boldsymbol{p}_{\boldsymbol{k},t}\}$; and (2), we let the tolerance $\epsilon\rightarrow0$. We then have
\begin{align*}
&P_\text{SFABC}(\boldsymbol{k}| \{X_{t}\}_{t\in\mathcal{T}}) 
\nonumber\\&  \xrightarrow{\mathsf{(1)}} \int_{\boldsymbol{\mathcal{P}}} \mathrm{d}\{\boldsymbol{p}'_t\}_{t\in\mathcal{T}} \frac{\mathbb{I}_{\boldsymbol{\mathcal{P}}_{\boldsymbol{k}}^{\boldsymbol{\mathcal{M}}}}(\max_{t\in\mathcal{T}} d_{\boldsymbol{\mathcal{U}}_t}(\boldsymbol{p}_{\boldsymbol{k},t},\boldsymbol{p}'_t)\leq \epsilon)}{V_{\epsilon}^{\boldsymbol{\mathcal{X}}}(\boldsymbol{k})}  \times \nonumber\\ & \quad \quad \quad \ \prod_{t\in\mathcal{T}}\prod_{b=1}^{B_t}  p'_t(\boldsymbol{u}_{b,t})^{n_{b,t}} \ \pi(\boldsymbol{k}) \nonumber\\
&\xrightarrow{\mathsf{(2)}} \int_{\boldsymbol{\mathcal{P}}} 
\mathrm{d}\{\boldsymbol{p}'_t\}_{t\in\mathcal{T}} \prod_{t\in\mathcal{T}} \delta(\boldsymbol{p}_{\boldsymbol{k},t},\boldsymbol{p}'_t) \prod_{t\in\mathcal{T}}\prod_{b=1}^{B_t} p'_t(\boldsymbol{u}_{b,t})^{n_{b,t}} \ \pi(\boldsymbol{k})\nonumber\\
&= \prod_{t\in\mathcal{T}}\prod_{b=1}^{B_t} p_{\boldsymbol{k},t}(\boldsymbol{u}_{b,t})^{n_{b,t}} \pi(\boldsymbol{k}) \propto P(\boldsymbol{k} | \{X_{t}\}_{t\in\mathcal{T}}),
\end{align*}
where the last line is the posterior in \eqref{eqn:posterior}, which completes the proof.

\section*{Author Contributions}
Author contributions: Z.L., M.B and P.T. designed the research, interpreted results and wrote the paper; Z.L. performed the derivations, designed the computations and analyzed the data.

\begin{acknowledgments}
ZL acknowledges support from the Roth Scholarship awarded by the Department of Mathematics at Imperial College London; MB is supported by EPSRC through the EPSRC Centre for Mathematics of Precision Healthcare (EP/N014529/1); PT is supported by UKRI through a Future Leaders Fellowship (MR/T018429/1 and MR/Y034309/1).
\end{acknowledgments}

%Use \bibsplit to split the references from the body of the text. Value "[3]" represents the number of reference in the left column (Note: Please avoid single column figures & tables on this page.)

% Bibliography
\renewcommand{\href}[2]{#2}
\let\url\nolinkurl
\bibliography{bibliography}

\end{document}

\title[Supplementary Material]
{Supplementary Material for\\
Simulation-free approximate Bayesian computation
for stochastic reaction networks}

\author{Zekai Li}

\author{Mauricio Barahona}
\thanks{Corresponding authors:
m.barahona@imperial.ac.uk and p.thomas@imperial.ac.uk.}

\author{Philipp Thomas}
\thanks{Corresponding authors:
m.barahona@imperial.ac.uk and p.thomas@imperial.ac.uk.}

\affiliation{
Department of Mathematics, Imperial College London,
London SW7 2AZ, United Kingdom
}

\date{\today}

\maketitle

% Supplementary numbering
\renewcommand{\thesection}{S\arabic{section}}
\renewcommand{\theequation}{S\arabic{equation}}
\renewcommand{\thefigure}{S\arabic{figure}}
\renewcommand{\thetable}{S\arabic{table}}

% Your supplementary text starts here.
\section{Baseline ABC Algorithms}
The ABC rejection algorithm (ABC-Rej) \citep{wilkinson2013approximate, liepe2014framework} works as follows to approximate the posterior:
\vspace{2mm}
\begin{description}
    \item \textbf{Algorithm S1: ABC rejection sampling (ABC-Rej)} 
    \begin{enumerate} 
        \item[\textit{S1}] Sample a candidate parameter vector $\boldsymbol{k}$ from the prior $\pi(\cdot)$.
        \item[\textit{S2}]  Simulate datasets $\{X'_{t}\}_{t\in\mathcal{T}}$ from the model with the parameter vector $\boldsymbol{k}$ at observation times $\mathcal{T}$.
        \item[\textit{S3}] For a given tolerance $\epsilon$, accept $\boldsymbol{k}$ if $\max_{t\in\mathcal{T}} d(X'_t,X_t)\leq \epsilon$ with a distance function $d$. Otherwise, go to \textit{S1}.
    \end{enumerate}
\end{description}
\vspace{2mm}

ABC with sequential Monte Carlo (ABC-SMC) has been proposed to improve sampling efficiency by leveraging the ABC-Rej algorithm \citep{toni2009approximate}. To collect $N$ samples from the posterior with a desired tolerance $\epsilon$, the SFABC-SMC algorithm samples a series of populations of weighted samples, $\boldsymbol{\mathcal{K}}_l\coloneqq\{\boldsymbol{k}^{(i)}_{l}\}_{i=1}^{N}$ with weights $\boldsymbol{w}_{l} \coloneqq\{w^{(i)}_{l}\}_{i=1}^{N}$ for $l=1,\ldots,L$. Each population contains samples accepted with shrinking tolerances $\epsilon_1>\cdots>\epsilon_L=\epsilon$, where $\{\epsilon_l\}_{l=1}^{L}$ is a chosen tolerance sequence. 
Given a tolerance sequence $\{\epsilon_l\}_{l=1}^{L}$ and perturbation kernels $\{\Psi_l\}_{l=1}^{L}$, the ABC-SMC algorithm proceeds as follows:
\vspace{2mm}
\begin{description} 
    \item \textbf{Algorithm S2: ABC with sequential Monte Carlo sampling (ABC-SMC)} 
    \begin{enumerate} 
        \item[\textit{S1}] Set tolerances $\epsilon_1, \ldots, \epsilon_L$. Set population indicator $l=1$.
        \item[\textit{S2.0}] Set sample count $i=1$.
        \item[\textit{S2.1}] If $l=1$, sample $\boldsymbol{k}$ from the prior $\pi(\cdot)$. \\ Else, sample $\boldsymbol{k}'$ from the previous population $\boldsymbol{\mathcal{K}}_{l-1}$ with weights $\boldsymbol{w}_{l-1}$. Perturb $\boldsymbol{k}'$ to obtain $\boldsymbol{k}\sim \Psi_l(\cdot|\boldsymbol{k}')$ where $\Psi_l$ is a perturbation kernel. If $\pi(\boldsymbol{k})=0$, return to \textit{S2.1}.
        \item[\textit{S2.2}] Simulate datasets $\{X'_{t}\}_{t\in\mathcal{T}}$ from the model with the parameter vector $\boldsymbol{k}$ at observation times $\mathcal{T}$. For tolerance $\epsilon_l$, accept $\boldsymbol{k}$ if $\max_{t\in\mathcal{T}^*} d(X'_t,X_t)\leq \epsilon_l$ with a distance function $d$. Otherwise, go to \textit{S2.1}.
        \item[\textit{S2.3}] If $i<N$, set $i = i+1$, return to \textit{S2.1}. \\ Else, calculate the weight of each sample $\boldsymbol{k}^{(i)}_{l}$: 
        $$w^{(i)}_l = \begin{cases}
              1, & \text{if}\ l=1 \\
              \frac{\pi(\boldsymbol{k}^{(i)}_{l})}{\sum_{j=1}^N w^{(j)}_{l-1}\Psi_l\left(\boldsymbol{k}^{(i)}_{l}|\boldsymbol{k}^{(j)}_{l-1}\right)}, & \text{if}\ l>1
            \end{cases} $$
        \item[\textit{S3}] If $l<L$, set $l = l+1$ and go to \textit{S2.0}.
    \end{enumerate}
\end{description}
\vspace{2mm}

\section{Feasibility Test in SFABC}
One of the core ideas of SFABC is the feasibility of the sampled parameter set and the sampled candidate state probability. In practice, this feasibility problem is often intractable because evaluating the state probability $\boldsymbol{p}_{\boldsymbol{k},t}$ requires solving the CME. This feasibility problem can be formulated as finding a feasible solution within the sets specified by the CME constraint and other constraints, but these sets may be infinite-dimensional due to an unbounded state space. Therefore, we relax the constraints in the set to obtain a solvable formulation.

The first approximation is to truncate the state space. We write the following CME over truncated state space $\boldsymbol{\mathcal{M}}\subseteq\boldsymbol{\mathcal{X}}$:
\begin{equation} \label{eqn:CME}
    \frac{\mathrm{d} \boldsymbol{p}^{\boldsymbol{\Tilde{\mathcal{M}}}}_{\boldsymbol{k},t}}{\mathrm{d} t} = \sum_{r=1}^R k_r Q_r \boldsymbol{p}^{\boldsymbol{\mathcal{M}}}_{\boldsymbol{k},t}
\end{equation} 
for $\boldsymbol{p}^{\boldsymbol{\Tilde{\mathcal{M}}}}_{\boldsymbol{k},t}\coloneqq(P_t(\boldsymbol{x}|\boldsymbol{k}))_{\boldsymbol{x}\in\boldsymbol{\Tilde{\mathcal{M}}}}$ and $\boldsymbol{p}_{\boldsymbol{k},t}\coloneqq(P_t(\boldsymbol{x}|\boldsymbol{k}))_{\boldsymbol{x}\in\boldsymbol{\mathcal{M}}}$ where $\boldsymbol{\Tilde{\mathcal{M}}}$ is the maximal subset of $\boldsymbol{\mathcal{M}}$ which does not depend on states outside $\boldsymbol{\mathcal{M}}$ \citep{kuntz2019bounding}. Here, the coefficient matrix $Q_r \in \mathcal{R}^{|\boldsymbol{\Tilde{\mathcal{M}}}|\times|\boldsymbol{\mathcal{M}}|}$ has entries $Q_r(\boldsymbol{x},\boldsymbol{y}) = a_r(\boldsymbol{y})(\mathbb{I}_{\boldsymbol{y}+\boldsymbol{v}_r}(\boldsymbol{x})-\mathbb{I}_{\boldsymbol{y}}(\boldsymbol{x}))$ for $\boldsymbol{x}\in\boldsymbol{\Tilde{\mathcal{M}}}$ and $\boldsymbol{y} \in \boldsymbol{\mathcal{M}}$. In the rest of this section, we simplify the notation and write $\boldsymbol{\Tilde{p}}_{\boldsymbol{k},t}\coloneqq\boldsymbol{p}^{\boldsymbol{\Tilde{\mathcal{M}}}}_{\boldsymbol{k},t}$ and $\boldsymbol{p}_{\boldsymbol{k},t}\coloneqq\boldsymbol{p}^{\boldsymbol{\mathcal{M}}}_{\boldsymbol{k},t}$ for the truncated solutions to the CME with parameter $\boldsymbol{k}$, and denote by $\boldsymbol{\Tilde{p}}_t$ and $\boldsymbol{p}_t$ a general probability vector over at time $t$ subsets $\boldsymbol{\Tilde{\mathcal{M}}}$ and $\boldsymbol{\mathcal{M}}$, respectively. We assume the projection of $\mathcal{M}$ onto the observed species contains $\boldsymbol{\mathcal{U}}_t$ for all $t\in\mathcal{T}$

At the steady state, the dynamics of the network no longer evolve in time, and we have
\begin{equation} \label{eqn:CME-ss}
    \frac{\mathrm{d} \boldsymbol{\Tilde{p}}_{\boldsymbol{k}}}{\mathrm{d} t} = \sum_{r=1}^R k_r Q_r \boldsymbol{p}_{\boldsymbol{k}} = 0,
\end{equation} 
where the time subscripts on the probabilities are omitted. We assess the compatibility of the rate parameters $\boldsymbol{k}$ with a sampled state probability $\boldsymbol{p}'\sim \mathrm{Dir}(n_1+1,\ldots,n_B+1, 1)$ by checking whether a feasible solution $\boldsymbol{p}$ exists in the following set:
\begin{equation} \label{set:ss}
	\boldsymbol{\mathcal{P}}_{\boldsymbol{k}}^{\boldsymbol{\mathcal{M}}} = \left\{{\setstretch{1.314} \begin{array}{l}\boldsymbol{p}\in\mathbb{R}_+^{|\boldsymbol{\mathcal{M}}|\times1}\end{array}} {\setstretch{1.5} \begin{array}{|ll}
            \sum_{\boldsymbol{x}\in\boldsymbol{\mathcal{M}}} \boldsymbol{p}(\boldsymbol{x}) \leq 1\\
			\sum_{r=1}^R k_r Q_r \boldsymbol{p} = 0 \\
			d_{\boldsymbol{\mathcal{U}}}(\boldsymbol{p},\boldsymbol{p}') \leq \epsilon
	\end{array}}\right \},
\end{equation}
where $\boldsymbol{\mathcal{U}}$ is the set of unique states in the observed data and the distance function $d_{\boldsymbol{\mathcal{U}}}(\boldsymbol{p},\boldsymbol{p}')$ compares the probabilities over $\boldsymbol{\mathcal{U}}$ (potentially only over observable species $\mathcal{S}$).

In contrast to the steady-state case with the exact dynamical constraint, the CME constraints involve integrals over the unobserved trajectories between observation times, so they cannot be enforced exactly from snapshots alone. To link the rate parameters with snapshot observations, we multiply an exponential test function $e^{-\rho t}$ to both sides of the CME for a given exponential weighting coefficient $\rho$ and integrate by parts from $t_1$ to $t_2$ with $t_2> t_1\geq 0$, which yields:
\begin{equation} \label{eqn:CME-tran}
    e^{-\rho t_2}\boldsymbol{\Tilde{p}}_{\boldsymbol{k},t_2} - e^{-\rho t_1}\boldsymbol{\Tilde{p}}_{\boldsymbol{k},t_1} + \rho\int_{t_1}^{t_2} e^{-\rho t}\boldsymbol{\Tilde{p}}_{\boldsymbol{k},t} \mathrm{d}t = \sum_{r=1}^R k_rQ_r \int_{t_1}^{t_2} e^{-\rho t}\boldsymbol{p}_{\boldsymbol{k},t} \mathrm{d}t,
\end{equation} 
for a given coefficient $\rho\in \mathbb{R}$. Defining $\boldsymbol{q}_{\boldsymbol{k},t_1,t_2;\rho} \coloneqq \int_{t_1}^{t_2} e^{-\rho t}\boldsymbol{p}_{\boldsymbol{k},t} \mathrm{d}t$ and $\boldsymbol{\Tilde{q}}_{\boldsymbol{k},t_1,t_2;\rho}\coloneqq \int_{t_1}^{t_2} e^{-\rho t}\boldsymbol{\Tilde{p}}_{\boldsymbol{k},t} \mathrm{d}t$, we can rewrite \eqref{eqn:CME-tran} as
\begin{equation} \label{eqn:CME-tran-simple}
    e^{-\rho t_2}\boldsymbol{\Tilde{p}}_{\boldsymbol{k},t_2} - e^{-\rho t_1}\boldsymbol{\Tilde{p}}_{\boldsymbol{k},t_1} + \rho\boldsymbol{\Tilde{q}}_{\boldsymbol{k},t_1,t_2;\rho}= \sum_{r=1}^R k_rQ_r \boldsymbol{q}_{\boldsymbol{k},t_1,t_2;\rho}.
\end{equation} 
With observations at time points $\mathcal{T}$, we consider time points $\mathcal{T}_{\text{opt}} = \{t_1,\ldots,t_{T'}\}$ with $0\leq t_1< \ldots< t_{T'}=t_{\text{max}}$ in the optimization, where $t_{\text{max}}$ denotes the largest observation time and we assume $\mathcal{T}\subseteq\mathcal{T}_{\text{opt}}$, and we denote $\boldsymbol{q}_{i;\rho}\coloneqq\int_{t_i}^{t_{i+1}} e^{-\rho t}\boldsymbol{p}_{t} \mathrm{d}t$ and $\boldsymbol{\Tilde{q}}_{i;\rho}\coloneqq\int_{t_i}^{t_{i+1}} e^{-\rho t}\boldsymbol{\Tilde{p}}_{t} \mathrm{d}t$ for $i=1,\ldots,T'-1$. For a chosen set of exponential weighting coefficients $\rho \in \mathcal{R}$ in which we assume $\rho = 0$ is contained, we can write the constrained set for the transient case as:
\begin{equation} \label{set:tran}
	\boldsymbol{\mathcal{P}}_{\boldsymbol{k}}^{\boldsymbol{\mathcal{M}}} = \left\{{\setstretch{1.314} \begin{array}{l}\boldsymbol{p}_t\in\mathbb{R}_+^{|\boldsymbol{\mathcal{M}}|\times1}\text{ for }t \in \mathcal{T}_{\text{opt}} \\ \boldsymbol{q}_{i;\rho} \in\mathbb{R}_+^{|\boldsymbol{\mathcal{M}}|\times1}\\\text{for } i=1,\ldots,T'-1\\ \text{and } \rho\in\mathcal{R} \end{array}} {\setstretch{1.5} \begin{array}{|ll}
            \sum \boldsymbol{p}_t \leq 1, & t\in\mathcal{T}_{\text{opt}}\\
            \sum \boldsymbol{q}_{i;0} \leq t_{i+1} - t_i, & i=1,\ldots,T-1\\
            \sum \boldsymbol{q}_{i;\rho} \leq (e^{-\rho t_i} - e^{-\rho t_{i+1}})/\rho, & i=1,\ldots,T-1;\rho\neq0\in\mathcal{R}\\
            e^{-(\rho_1-\rho_2)t_{i+1}}\boldsymbol{q}_{i;\rho_2}\leq \boldsymbol{q}_{i;\rho_1}\leq e^{-(\rho_1-\rho_2)t_i}\boldsymbol{q}_{i;\rho_2}, & i=1,\ldots,T-1;\rho_1<\rho_2\in\mathcal{R}\\
			e^{-\rho t_{i+1}}\boldsymbol{\Tilde{p}}_{t_{i+1}} - e^{-\rho t_i}\boldsymbol{\Tilde{p}}_{t_i} + \rho\boldsymbol{\Tilde{q}}_{i;\rho}= \sum_{r=1}^R k_rQ_r \boldsymbol{q}_{i;\rho}, & i=1,\ldots,T-1;\rho\in\mathcal{R} \\
			d_{\boldsymbol{\mathcal{U}}_t}(\boldsymbol{p}_t,\boldsymbol{p}'_t) \leq \epsilon, & t\in\mathcal{T} \\
	\end{array}}\right \},
\end{equation}
where the state probability $\boldsymbol{p}'_t$ is obtained as a Dirichlet resampling of the data at time $t\in\mathcal{T}$. Here, the first four constraints ensure that the probability and integral variables are valid and properly normalized, while the fifth one is the CME constraint and the last one is the distance constraint. Since the truncated transient CME is a finite-dimensional linear ODE in time, enforcing the CME-derived constraints only on the intervals of $\mathcal{T}_{\text{opt}}$ is an ordinary time discretization, and as the mesh size $\max_i(t_{i+1,\text{opt}}-t_{i,\text{opt}})\rightarrow0$ the discrete constraints converge to the continuous-time constraint.

Model-specific constraints can also be incorporated into the feasibility check, such as a steady-state constraint on the initial total-mRNA distribution in the metabolic-labeling model or a binomial partitioning constraint relating the mRNA-count distributions immediately before and after cell division. Additionally, \eqref{eqn:CME-tran-simple} also holds for piecewise constant rate parameters, as we can include all times at which any parameter value changes in the set $\mathcal{T}_{\text{opt}}$.

\section{SFABC with Sequential Monte Carlo}

To improve sampling efficiency, we propose SFABC with a sequential Monte Carlo (SFABC-SMC) sampling scheme. Given a tolerance sequence $\{\epsilon_l\}_{l=1}^{L}$ and perturbation kernels $\{\Psi_l\}_{l=1}^{L}$, the SFABC-SMC proceeds as follows:
\vspace{2mm}
\begin{description} 
    \item \textbf{Algorithm S3: SFABC sequential Monte Carlo (SFABC-SMC)} 
    \begin{enumerate} 
        \item[\textit{S1}] Set tolerances $\epsilon_1, \ldots, \epsilon_L$. Set population indicator $l=1$.
        \item[\textit{S2.0}] Set sample count $i=1$.
        \item[\textit{S2.1}] If $l=1$, sample $\boldsymbol{k}$ from the prior $\pi(\cdot)$. \\ Else, sample $\boldsymbol{k}'$ from the previous population $\boldsymbol{\mathcal{K}}_{l-1}$ with weights $\boldsymbol{w}_{l-1}$. Perturb $\boldsymbol{k}'$ to obtain $\boldsymbol{k}\sim \Psi_l(\cdot|\boldsymbol{k}')$ where $\Psi_l$ is a perturbation kernel. If $\pi(\boldsymbol{k})=0$, return to \textit{S2.1}.
        \item[\textit{S2.2}] For each observation time $t\in\mathcal{T}$, sample a candidate state probability $\boldsymbol{p}'_t\sim\mathrm{Dir}(\boldsymbol{\alpha}(X_t))$.
        \item[\textit{S2.3}]  Test if there exists a CME relaxation $\{\boldsymbol{p}_t\}_{t\in\mathcal{T}}\in\boldsymbol{\mathcal{P}}_{\boldsymbol{k}}^{\boldsymbol{\mathcal{M}}}$ satisfies $\max_{t\in\mathcal{T}} d_{\boldsymbol{\mathcal{U}}_t}(\boldsymbol{p}_{t},\boldsymbol{p}'_t)\leq\epsilon_l$. If such a solution exists, accept $\boldsymbol{k}$. Otherwise, go to \textit{S2.1}.
        \item[\textit{S2.4}] If $i<N$, set $i = i+1$, return to \textit{S2.1}. \\ Else, calculate the volume $V_{\epsilon_l}(\boldsymbol{k}^{(i)}_{l})$ (see \textit{Materials and Methods} in the main text). Calculate the weight of each sample $\boldsymbol{k}^{(i)}_{l}$: 
        $$w^{(i)}_l = \begin{cases}
              1/V_{\epsilon_l}(\boldsymbol{k}^{(i)}_{l}), & \text{if}\ l=1 \\
              \frac{\pi(\boldsymbol{k}^{(i)}_{l})}{V_{\epsilon_l}(\boldsymbol{k}^{(i)}_{l})\sum_{j=1}^N w^{(j)}_{l-1}\Psi_l\left(\boldsymbol{k}^{(i)}_{l}|\boldsymbol{k}^{(j)}_{l-1}\right)}, & \text{if}\ l>1
            \end{cases} $$
        \item[\textit{S3}] If $l<L$, set $l = l+1$ and go to \textit{S2.0}.
    \end{enumerate}
\end{description}
\vspace{2mm}
The weighted population $\boldsymbol{w}_L\boldsymbol{\mathcal{K}}_L$ can be shown to approximate the distribution of accepted samples from an SFABC-Rej algorithm with tolerance $\epsilon_L=\epsilon$ in the way that the ABC-SMC algorithm approximates the ABC-Rej algorithm in \citep{toni2009approximate}.

The SFABC-SMC algorithm for parameter inference can be easily extended to model selection problems. Consider candidate models indexed by $\{1,\ldots,M\}$ and the model-specific parameters $\boldsymbol{k}(m)$ for $m=1,\ldots,M$, we use $\phi(\cdot)$ to denote the prior over these models and use $\pi_m(\cdot)$ to denote the prior of rate parameters given model $m$. Given tolerance sequence $\{\epsilon_l\}_{l=1}^{L}$, each population may contain samples from any model, namely $\boldsymbol{\mathcal{K}}_l\coloneqq\{\boldsymbol{k}^{(i)}_{l}\}_{i=1}^{N}$ with models $\boldsymbol{m}_l\coloneqq\{m^{(i)}_{l}\}_{i=1}^{N}$ and weights $\boldsymbol{w}_{l} \coloneqq\{w^{(i)}_{l}\}_{i=1}^{N}$ for $l=1,\ldots,L$, where each $m^{(i)}_{l}\in\{1,\ldots,M\}$. The SFABC-SMC model selection algorithm works as follows:
\vspace{2mm}
\begin{description} 
    \item \textbf{Algorithm S4: SFABC-SMC for Model Selection} 
    \begin{enumerate} 
        \item[\textit{S1}] Set tolerances $\epsilon_1, \ldots, \epsilon_L$. Set population indicator $l=1$.
        \item[\textit{S2.0}] Set sample count $i=1$.
        \item[\textit{S2.1}] Sample a model $m$ from $\phi(\cdot)$. \\ If $l=1$,  Sample $\boldsymbol{k}$ from the prior $\pi_m(\cdot)$. \\ Else, sample $\boldsymbol{k}'$ from the previous population $\boldsymbol{\mathcal{K}}_{l-1}^m$ with weights $\boldsymbol{w}_{l-1}^m$. Perturb $\boldsymbol{k}'$ to obtain $\boldsymbol{k}\sim \Psi_{l}^m(\cdot|\boldsymbol{k}'(m))$ where $\Psi_{l}^m$ is a perturbation kernel for model $m$. If $\pi_m(\boldsymbol{k})=0$, return to \textit{S2.1}.
        \item[\textit{S2.2}] For each observation time $t\in\mathcal{T}$, sample a candidate state probability $\boldsymbol{p}'_t\sim\mathrm{Dir}(\boldsymbol{\alpha}(X_t))$.
        \item[\textit{S2.3}]   Test if there exists a CME relaxation $\{\boldsymbol{p}_t\}_{t\in\mathcal{T}}\in\boldsymbol{\mathcal{P}}_{\boldsymbol{k}}^{\boldsymbol{\mathcal{M}}}$ satisfies $\max_{t\in\mathcal{T}} d_{\boldsymbol{\mathcal{U}}_t}(\boldsymbol{p}_{t},\boldsymbol{p}'_t)\leq\epsilon_l$. If such a solution exists, accept $\boldsymbol{k}$. Otherwise, go to \textit{S2.1}.
        \item[\textit{S2.4}] If $i<N$, set $i = i+1$, return to \textit{S2.1}. \\ Else, calculate the volumes $V_{\epsilon_l}(\boldsymbol{k}^{(i)}_{l})$. Calculate the weight of each sample $\boldsymbol{k}^{(i)}_{l}$: 
        $$w^{(i)}_l = \begin{cases}
              1/V_{\epsilon_l}(\boldsymbol{k}^{(i)}_{l}), & \text{if}\ l=1 \\
              \frac{\pi_m(\boldsymbol{k}^{(i)}_{l})}{V_{\epsilon_l}(\boldsymbol{k}^{(i)}_{l})\sum_{j:m^{(j)}_{l-1}=m} w^{(j)}_{l-1}\Psi^m_l\left(\boldsymbol{k}^{(i)}_{l}|\boldsymbol{k}^{(j)}_{l-1}\right)}, & \text{if}\ l>1
            \end{cases} $$
        \item[\textit{S3}] If $l<L$, set $l = l+1$ and go to \textit{S2.0}.
    \end{enumerate}
\end{description} 
\vspace{2mm}
With tolerance $\epsilon_L$, the posterior model probability is approximated as
\begin{equation} 
    P_\text{SFABC-SMC}(m|\{X_{t}\}_{t\in\mathcal{T}}) \approx \frac{1}{N}\sum_{i=1}^N \mathbb{I}(m^{(i)}_{L}=m),
\end{equation}
and the posterior distribution of parameters conditional on model $m$ is approximated by the weighted population $\{w^{(i)}_{L}\boldsymbol{k}^{(i)}_{L}\}_{i:m^{(i)}_{L}=m}$.

\section{Implementation Details of Inference Algorithms}

\subsection{Inference Setup} We summarize the inference setup for the ABC and SFABC algorithms in Table \ref{table:S1}, providing details on prior distribution, tolerance, type of constraints in the SFABC algorithms, and the choice of time points $\mathcal{T}$ and weightings $\mathcal{R}$ in the transient cases of SFABC. 

\subsection{Choice of SMC perturbation kernels}
In setups S2 and S4, the perturbation kernel used in the SMC samplers is
\begin{equation} \label{kernel:1}
    \Psi_l(\cdot|\boldsymbol{k}') \sim \text{Uniform}(k'-r_l,k'+r_l),
\end{equation}
independently for the three inferred parameters with $r_1 = 2.5$ and $r_2 = 1$. For setups S6 - S11, we use a perturbation kernel 
\begin{equation}
    \Psi(\cdot|\boldsymbol{k}') \sim k'\times\text{Uniform}(0.75,1.25)
\end{equation}
independently for each parameter. For setup S12, we use the following perturbation kernel:
\begin{equation}
    \Psi_l(\cdot|\boldsymbol{k}') \sim 10^{\text{Uniform}(\mathrm{log}_{\mathrm{10}}(k')-r,\mathrm{log}_{\mathrm{10}}(k')+r)}
\end{equation}
with $r_1=0.5$ and $r_2=0.25$, independently for each parameter in the current model.

\subsection{State space}
One also needs to choose the states in the optimization problem, namely the set $\boldsymbol{\mathcal{M}}$. For the telegraph model, we have $\boldsymbol{\mathcal{M}}=\{0,1\}\times\{0,\ldots,2x_{\text{max}}\}$, where $x_{\text{max}}$ is the largest observation. For the metabolic labeling model, we have $\boldsymbol{\mathcal{M}}=\{0,1\}\times\{0,\ldots,2u_{\text{max}}\}\times\{0,\ldots,2l_{\text{max}}\}$, where $u_{\text{max}},l_{\text{max}}$ are the largest observations of unlabeled and labeled mRNA counts, respectively. In the application to metabolic labeling data (S9 - S11 in Table \ref{table:S1}), the genes in D2 are ranked by the size of their joint state space, quantified as $\max(u)\times\max(l)$, and the 100 genes with the smallest size are selected.

\subsection{Summary statistics and distance functions}
We use standard moment-based distance functions. As the distances are calculated per time point, we describe the calculation for a single time point in this section and omit the time subscripts. For $\boldsymbol{x}=(x_1,\ldots, x_S)^\intercal$, we define  $\boldsymbol{x}^{\boldsymbol{\alpha}}\coloneqq \prod_{s=1}^S x_s^{\alpha_s}$ for $\boldsymbol{\alpha}\coloneqq(\alpha_1,\ldots, \alpha_S)^\intercal\in \mathbb{N}^S$ and denote the $\boldsymbol{\alpha}$-th moment of dataset $X=(\boldsymbol{x}_1,\ldots,\boldsymbol{x}_N)$ to be $\mu^{\boldsymbol{\alpha}}(X)\coloneqq\frac{1}{N}\sum_{i=1}^{N} \boldsymbol{x}_i^{\boldsymbol{\alpha}}$. 

For ABC, we use
\begin{equation}
    d(X',X) = \text{max}_{|\boldsymbol{\alpha}|\leq D}\left|\frac{\mu^{\boldsymbol{\alpha}}(X')-\mu^{\boldsymbol{\alpha}}(X)}{\hat{\text{se}}(\mu^{\boldsymbol{\alpha}}(X))}\right|,
\end{equation}
where $|\boldsymbol{\alpha}|\coloneqq\sum_{s=1}^S \alpha_s$ is the order of the moment and $\hat{\text{se}}(\mu^{\boldsymbol{\alpha}}(X))$ is the bootstrap estimate of the standard deviation of the sample moment with 1000 bootstrap samples. 

For SFABC, the candidate state probability is over the observed states with an additional bin representing the sum of probabilities over the unobserved states. Defining the moments over observed states $\mu_{\text{obs}}^{\boldsymbol{\alpha}}(\boldsymbol{p})\coloneqq\sum_{b=1}^B p(\boldsymbol{u}_b) {\boldsymbol{u}_b}^{\boldsymbol{\alpha}}$, we consider the distance function:
\begin{equation}
    d_{\boldsymbol{\mathcal{U}}}(\boldsymbol{p}^{\boldsymbol{\mathcal{M}}}_{\boldsymbol{k}},\boldsymbol{p}') = \text{max}_{|\boldsymbol{\alpha}|\leq D}\left|\frac{\mu_{\text{obs}}^{\boldsymbol{\alpha}}(\boldsymbol{p}^{\boldsymbol{\mathcal{M}}}_{\boldsymbol{k}}(\boldsymbol{\mathcal{U}}))-\mu_{\text{obs}}^{\boldsymbol{\alpha}}(\boldsymbol{p}'(\boldsymbol{\mathcal{U}}))}{\hat{\text{se}}(\mu_{\text{obs}}^{\boldsymbol{\alpha}}(\boldsymbol{p}'(\boldsymbol{\mathcal{U}})))}\right|.
\end{equation}
where $\boldsymbol{p}(\boldsymbol{\mathcal{U}})$ denotes the probability over the observed states $\boldsymbol{\mathcal{U}}$, and $\hat{\text{se}}(\mu_{\text{obs}}^{\boldsymbol{\alpha}}(\boldsymbol{p}'(\boldsymbol{\mathcal{U}})))$ is the standard deviation of the $\boldsymbol{\alpha}$-th moment estimated by 1000 samples from the Dirichlet distribution for $|\boldsymbol{\alpha}|>0$. A remarkable difference from ABC's distance function, in which $|\boldsymbol{\alpha}|=0$ leads to a trivial comparison, is that the case $|\boldsymbol{\alpha}|=0$ in SFABC compares the probability over the unobserved states, and it is necessary to include this in the distance function for an unbiased posterior estimate. In all numerical experiments, we set $\hat{\text{se}}(\mu_{\text{obs}}^{\boldsymbol{0}}(\boldsymbol{p}'(\boldsymbol{\mathcal{U}})))$ to be three standard deviations estimated from 1,000 samples from the Dirichlet distribution, where $\boldsymbol{0}\coloneqq(0,\ldots,0)^\intercal\in \mathbb{N}^S$. We choose $D=5$ for the telegraph model (S1 - S10, S12 in Table \ref{table:S1}) and $D=3$ for the metabolic labeling model (S11 in Table \ref{table:S1}) so that the total number of moments is the same.

\subsection{Exact MCMC Sampler for Steady-State Telegraph Model}
The steady-state probability of the telegraph model follows the Beta-Poisson distribution \citep{kim2013inferring, raj2006stochastic}, which can be used to evaluate the likelihood. For benchmarking, we implement the Metropolis-within-Gibbs algorithm, which provides exact samples from the posterior, and compare them with ABC and SFABC samples (S1-S4 in Table \ref{table:S1}). We run 10 independent chains for 55,000 iterations, discard the first 5,000 as burn-in, and keep every 1,000th draw thereafter (thinning). For each chain we initialize $k_{\text{off}}\sim\mathrm{Gamma}(3,1)$, $k_{\text{on}}\sim\mathrm{Gamma}(10,0.75)$, and $k_{\text{syn}}\sim\mathrm{Gamma}(5,1)$, and sample latent variables $p_j\sim\mathrm{Beta}(k_{\text{on}},k_{\text{off}})$ for $j=1,\ldots,N$. Each MCMC iteration updates the blocks $(\boldsymbol{p},k_{\text{on}},k_{\text{off}},k_{\text{syn}})$ sequentially using random-walk Metropolis steps. First, we propose $\boldsymbol{p}'=\boldsymbol{p}+\sigma_p\,\boldsymbol{\eta}$ with $\eta_j\sim\mathcal{N}(0,1)$ and only consider component-wise proposals that satisfy $0<p'_j<1$; valid components are accepted with the usual Metropolis rule based on the log-posterior difference. Next, we update $k_{\text{on}}$, $k_{\text{off}}$, and $k_{\text{syn}}$ in turn with proposals $k_m'=k_m+\sigma_{k_m} z$ (with $z\sim\mathcal{N}(0,1)$), rejecting immediately if $k_m'\le 0$ and otherwise accepting according to the corresponding log-posterior ratio. We aggregate the thinned samples across chains to form posterior draws of $\boldsymbol{k}=(k_{\text{off}},k_{\text{on}},k_{\text{syn}})$.

\section{List of Stochastic Reaction Networks} There are two stochastic reaction networks considered in this paper: the telegraph model and the metabolic labeling model. The telegraph model involves the following interactions:
\begin{equation} \label{model:telegraph}
    G_{\text{on}} \xrightleftharpoons[k_{\text{on}} g_{\text{off}}]{k_{\text{off}} g_{\text{on}}} G_{\text{off}}, \quad \varnothing \xrightarrow{k_{\text{syn}}g_{\text{on}}} X, \quad X \xrightarrow{k_{\text{deg}}x} \varnothing,
\end{equation}
where $G_{\text{on}}$ and $G_{\text{off}}$ denote the on and off states of a gene, and $X$ represents the mRNA count. We have $g_{\text{on}},g_{\text{off}}\in\{0,1\}$ with $g_{\text{on}}+g_{\text{off}}=1$, and $x\in\mathbb{N}$. Here, $k_{\text{off}}$ and $k_{\text{on}}$ are the switching rates between the on and off states, $k_{\text{syn}}$ is the synthesis rate of mRNA molecules, and $k_{\text{deg}}$ is the degradation rate. 

The metabolic labeling model, as an extension of the telegraph model, has the following interactions:
\begin{align} \label{model:meta}
    &G_{\text{on}} \xrightleftharpoons[k_{\text{on}} g_{\text{off}}]{k_{\text{off}} g_{\text{on}}} G_{\text{off}}, \quad \varnothing \xrightarrow{(1-\lambda) k_{\text{syn}}g_{\text{on}}} U, \quad \varnothing \xrightarrow{\lambda k_{\text{syn}}g_{\text{on}}} L, \quad U \xrightarrow{k_{\text{deg}}u} \varnothing, \quad L \xrightarrow{k_{\text{deg}}l} \varnothing,
\end{align}
where $U$ is the count of the unlabeled mRNAs and $L$ is the count of the labeled ones. The additional parameter $\lambda$ is the labeling efficiency.

\section{Filtering scRNA-seq Data}
For dataset D1 of mouse fibroblasts from \cite{larsson2019genomic}, we considered only the CAST allele and obtained the processed UMI counts from the accompanying repository \cite{sandberglab2019txburst}. First, we filtered out genes with more than 1\% missing counts. For the remaining genes, we filled missing counts with zero and further removed genes with a mean UMI count not exceeding 1 across cells. The remaining 979 genes were also observed in dataset D2 of mouse fibroblasts from \cite{ramskold2024single}. After inference with SFABC-SMC using setups S7 and S8 in Table \ref{table:S1}, we attempted to collect 100 posterior samples for each gene within the computational budget. Only genes for which 100 samples were obtained at the lowest tolerance were used for Fig. 3, leaving 764 genes for D1 and 865 genes for D2. We analyzed the total count (labeled and unlabeled) in Fig. 3. For Fig. 4, we used D2 and further selected 100 genes with the smallest product of the maximum UMIs across cells of the unlabeled ($u$) and 4sU-labeled ($l$) mRNAs, as the computational cost of SFABC scales with the size of the joint state space.

For the RPE-1 dataset with cell cycle reporters in \cite{battich2020sequencing}, we used the 2,364 cells under the pulse metabolic labeling condition and considered the total mRNA count. We first filtered the 11,848 genes to remove those with mean expression below one at all cell-cycle phases. For the remaining 2,987 genes, z-scores across the cell cycle were computed and clustered into three groups using $k$-means. Ten genes per cluster were then randomly sampled, yielding 30 genes for downstream SFABC analysis (gene names listed in Table \ref{tab:gene_clusters}).

\renewcommand{\href}[2]{#2}
\let\url\nolinkurl
\bibliography{bibliography}

\begin{table}\centering
\rotatebox{90}{
\begin{tabular}{|c|c|c|c|c|c|c|c|c|c|c|c|c|c|c|}
\hline
 \textbf{Setup}  & \textbf{Model} & \textbf{Prior} &  \textbf{Sampling Algorithm} & \textbf{Tolerance}  & \textbf{Constraints} & \textbf{Transient Optimization} \\
\hline
S1 &\multirow{4}{*}{Telegraph} & \multirow{4}{*}{\makecell[c]{$k_{\text{off}}\sim\text{Gamma}(3,1)$ \\
                 $k_{\text{on}}\sim\text{Gamma}(10,0.75)$ \\
                 $k_{\text{syn}}\sim\text{Gamma}(5,1)$}} & ABC-Rej & $2$  & - & - \\ \cline{1-1}\cline{4-7}

S2 & & & ABC-SMC & $\{10,5,2\}$ & - & -\\ \cline{1-1}\cline{4-7}

S3 & & &  SFABC-Rej & $2$  & Steady-state & - \\ \cline{1-1}\cline{4-7}

S4 & & &  SFABC-SMC & $\{10,5,2\}$ & Steady-state & - \\ \hline

S5 &\multirow{2}{*}{Telegraph} & \multirow{2}{*}{$k_{\text{off}},k_{\text{on}},k_{\text{syn}}\sim10^{\text{Unif}(-1,1)}$} & SFABC-SMC & $\{10,5,2\}$ & Steady-state & -\\ \cline{1-1}\cline{4-7}

S6 & & & ABC-SMC & $\{10,5,2\}$ & - & -   \\ \hline

S7 &\multirow{2}{*}{Telegraph} & \multirow{2}{*}{\makecell[c]{$k_{\text{off}},k_{\text{on}}\sim10^{\text{Unif}(-1.5,1.5)}$ \\
                 $k_{\text{syn}}\sim10^{\text{Unif}(-1.5,3)}$}} & SFABC-SMC & $\{10,6,4\}$ & Steady-state & -  \\ \cline{1-1}\cline{4-7}

S8 & & &  SFABC-SMC & $\{50,20,10\}$ & Steady-state & - \\ \hline

S9 & Telegraph & \makecell[c]{$k_{\text{off}},k_{\text{on}},k_{\text{deg}}\sim10^{\text{Unif}(-1.5,1.5)}$ \\
                 $k_{\text{syn}}\sim10^{\text{Unif}(-1.5,3)}$} &SFABC-SMC & $\{50,20,10\}$ & Steady-state & - \\ \hline
S10 & Telegraph &  \makecell[c]{$k_{\text{off}},k_{\text{on}},k_{\text{deg}}\sim10^{\text{Unif}(-1.5,1.5)}$ \\
                 $\lambda k_{\text{syn}}\sim10^{\text{Unif}(-1.5,3)}\cdot10^{\text{Unif}(-1,0)}$} &  SFABC-SMC & $\{50,20,10\}$ &  Transient & \makecell[c]{$\mathcal{T}=\{0,1,2\}$\\$|\mathcal{R}|=4$} \\ \hline
S11 & \makecell[c]{Metabolic\\labeling}& \makecell[c]{$k_{\text{off}},k_{\text{on}},k_{\text{deg}}\sim10^{\text{Unif}(-1.5,1.5)}$ \\
                 $k_{\text{syn}}\sim10^{\text{Unif}(-1.5,3)}$\\$\lambda\sim 10^{\text{Unif}(-1,0)}$}&  SFABC-SMC & $\{50,20,10\}$ & \makecell[c]{Steady-state$^\dagger$\\\& transient} & \makecell[c]{$\mathcal{T}=\{0,1,2\}$\\$|\mathcal{R}|=6$} \\ \hline

S12 & \makecell[c]{Cell-cycle\\telegraph}& \makecell[c]{$k_{\text{off}},k_{\text{on}}\sim10^{\text{Unif}(-3,3)}$ \\
                 $k_{\text{syn}}\sim10^{\text{Unif}(-2,2)}$\\$k_{\text{deg}}\sim 10^{\text{Unif}(-3,1.5)}$} &  SFABC-SMC & $\{10,7.5,5\}$ & \makecell[c]{Transient\\\& cell-cycle$^{\dagger\dagger}$} & \makecell[c]{$\mathcal{T}=\{0,0.5,1,\ldots,20\}$\\$|\mathcal{R}|=4$} \\ \hline
\end{tabular}
}
\caption{\textbf{Summary of Inference Setup.} For S1 - S8, we fix $k_{\text{deg}}\equiv1$. $\dagger$ The total mRNA count (i.e., labeled + unlabeled) is required to follow a steady-state distribution at $t=0\text{h}$ as the initial condition. $\dagger\dagger$ The cell-cycle telegraph model has a specific constraint forcing the distribution of mRNA count at $t=0\text{h}$ to follow a Binomial down-sampling with probability parameter 0.5 from that at $t=20\text{h}$.}
\label{table:S1}
\end{table}

\newpage
~
\begin{table}[ht]
\centering
\begin{tabular}{|l|l|l|}
\hline
\textbf{Cluster 1} & \textbf{Cluster 2} & \textbf{Cluster 3} \\
\hline
CENPB     & VEGFB     & KPNA2 \\
LTA4H     & \textbf{MEST}      & SPC25 \\
\textbf{GAK}       & MYCBP2    & CENPF \\
PDZD11    & ADAMTSL1  & \textbf{YBX3} \\
PLEKHA5   & FBN1      & EVA1A \\
TXNL4A    & CHAF1A    & ARHGEF2 \\
FH        & CTPS1     & PSMC3 \\
RPL4      & ARHGAP10  & RASSF3 \\
PABPC1    & SLC2A4RG  & ALYREF \\
RPL5P17   & SSH1      & CDCA8 \\
\hline
\end{tabular}
\caption{\textbf{Genes Used for SFABC Analysis in the Cell-Cycle Dataset \citep{battich2020sequencing}.} Related to Fig. 5. Genes were first filtered by mean expression, clustered into three groups based on their z-scored expression profiles across cell-cycle phases, and then ten genes were sampled from each cluster. Example genes in Fig. 5B are highlighted.}\label{tab:gene_clusters}
\end{table}

\newpage
~
\begin{figure}
\centering
\includegraphics{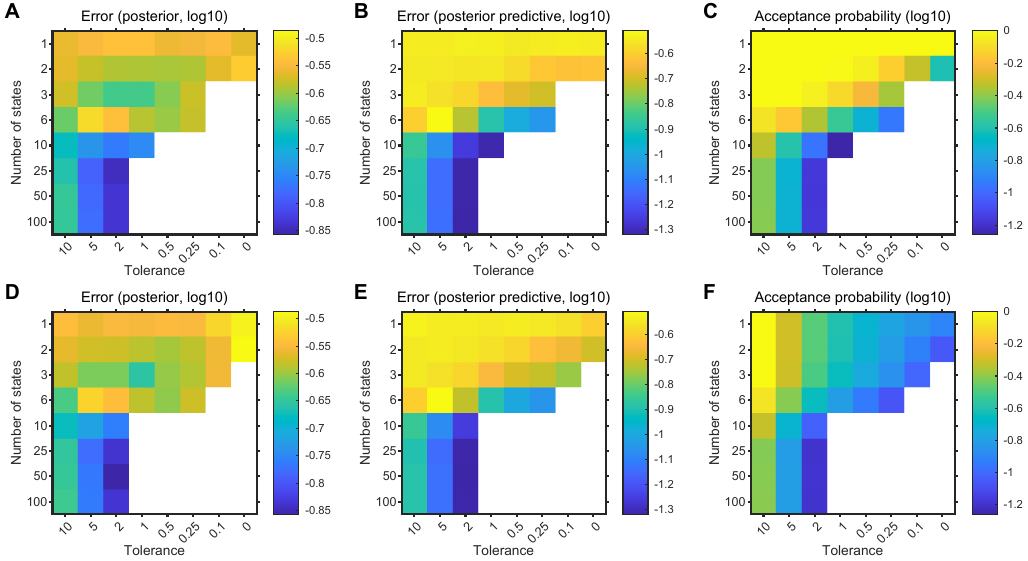}
\caption{\textbf{Effect of Truncated State-Space Size and Tolerance on SFABC Performance for Steady-State Telegraph-Model Data.} Heatmaps show validation metrics for the 20 synthetic datasets with sample size 500 used in Fig. 2B,C. The upper row shows SFABC-Rej (A--C), and the lower row shows SFABC-SMC (D--F). Columns show the posterior error relative to MCMC, the posterior predictive error relative to the observed data, and the acceptance probability, all on the $\log_{10}$ scale. Results are averaged over the 20 datasets for different tolerances and numbers of states included in the feasibility test. Empty cells indicate settings for which the required number of posterior samples was not obtained under the fixed optimization budgets: 15,000 optimizations for each SFABC-Rej run, and 25,000 optimizations for each SFABC-SMC run. While we expect the acceptance probability to decrease both as the tolerance is tightened and as the truncated state space is enlarged, small nonmonotonic fluctuations can occur because independent SFABC-Rej runs across different tolerances and truncation sizes used different random seeds. Except for the tolerance settings and truncated state-space sizes, the inference setup is the same as in S3-S4 in Table \ref{table:S1}. In SMC, we use the perturbation kernel in \eqref{kernel:1} with linearly decreased kernel parameter $r=\{2.5, 2.125, 1.75, 1.375, 1, 0.625, 0.25\}$.}
\label{fig:SuppFig0}
\end{figure}

\newpage
\begin{figure}
\centering
\includegraphics{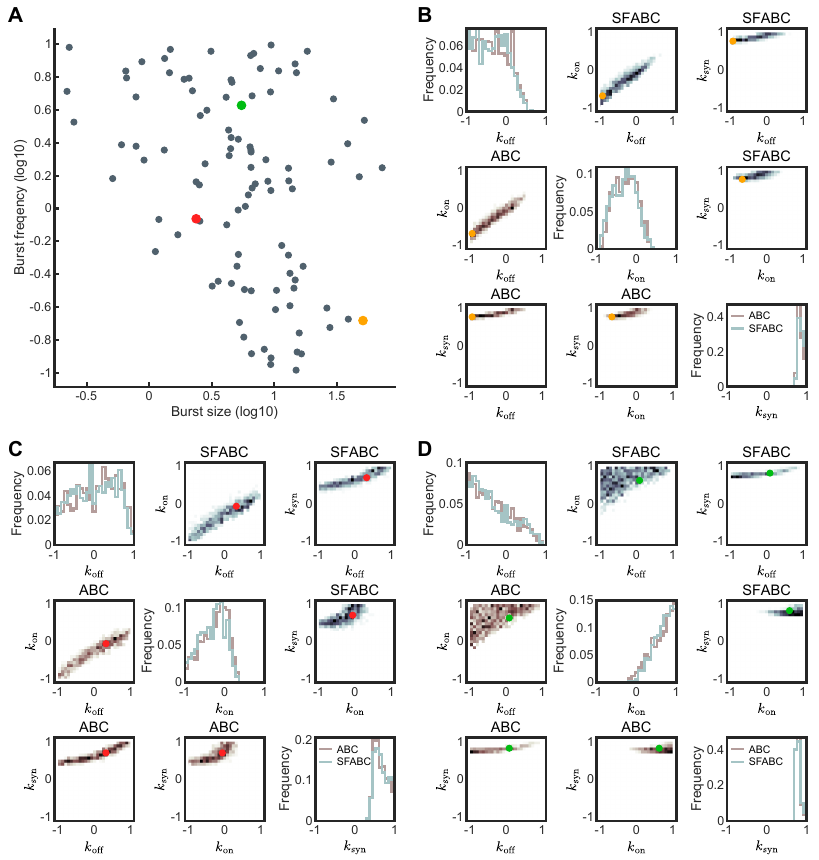}
\caption{\textbf{Comparison of ABC and SFABC Posterior Estimates for different synthetic data sets.} \textbf{(A)} Burst frequency and burst size of the 100 parameter sets used in the robustness test of SFABC in Fig 2E-H. Colored points represent the three example parameter sets in (B)-(D). \textbf{(B)-(D)} Posterior comparison between ABC and SFABC estimates. Diagonal sub-panels show marginal posterior distributions of the three parameters, and the other sub-panels compare the pairwise posterior distributions between them. In B-D, distributions are shown on a $\log_{10}$ scale.}\label{fig:SuppFig1}
\end{figure}

\newpage
\begin{figure}
\centering
\includegraphics{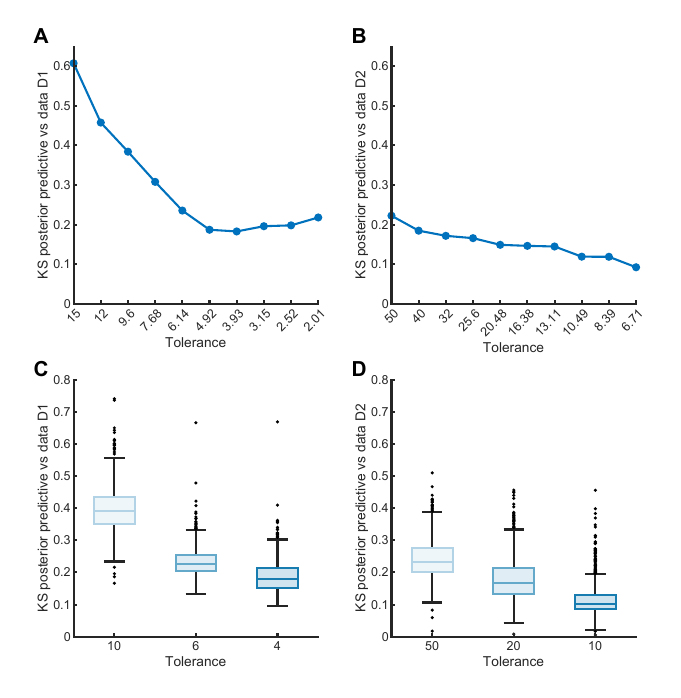}
\caption{\textbf{Selection of Tolerances and Validation of Posterior Estimates for Inference of the Telegraph Model with Steady-State scRNA-seq Datasets.} \textbf{(A)-(B)} Tolerance selection based on the Kolmogorov-Smirnov (KS) metrics. We test across ten tolerances for each dataset with SFABC-SMC, $\{15,15\times0.8^1,\ldots,15\times0.8^9\}$ for D1 and $\{50,50\times0.8^1,\ldots,50\times0.8^9\}$ for D2. The tolerance sequences selected for analysis in the main text are $\{10,6,4\}$ for D1 and $\{50,20,10\}$ for D2, balancing accuracy and computational cost. \textbf{(C)-(D)} KS metrics between SFABC's posterior predictives and observed datasets.}
\end{figure}

\newpage
\begin{figure}
\centering
\includegraphics{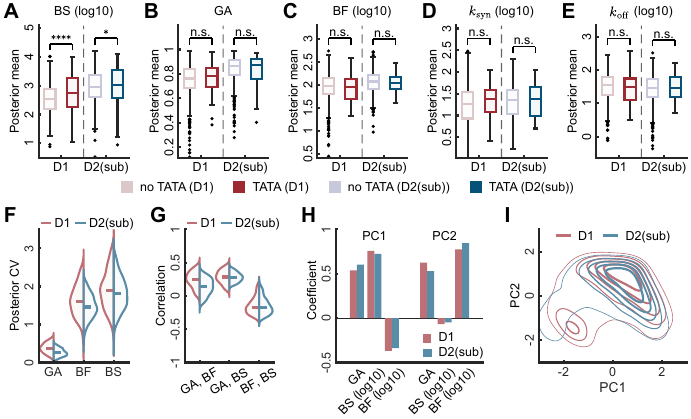}
\caption{\textbf{Ablation Study on the Larger Steady-State scRNA-seq Datasets.} Here, we randomly subsample the dataset D2 in \cite{ramskold2024single} with 8,916 cells to obtain D2(sub) with 224 cells, the same as in D1 in \cite{larsson2019genomic}. \textbf{(A)-(E)} Posterior means of burst kinetics and individual rate parameters, grouped by dataset and gene type (with or without a TATA element). \textbf{(F)-(G)} Coefficients of variation (CVs) of marginal posterior distributions and pairwise correlations in posterior samples across genes in each dataset, restricted to the 10th-90th percentiles. \textbf{(H)} Coefficients of principle components. For each gene, principal component analysis is performed on the (across-gene) posterior samples of GA, $\log_{10}(\text{BS})$, and $\log_{10}(\text{BF})$ for each of D1 and D2. \textbf{(I)} Distribution of posterior samples projected to the plane of the first principal component (PC1) and the second one (PC2).}
\end{figure}

\newpage
\begin{figure}
\centering
\includegraphics{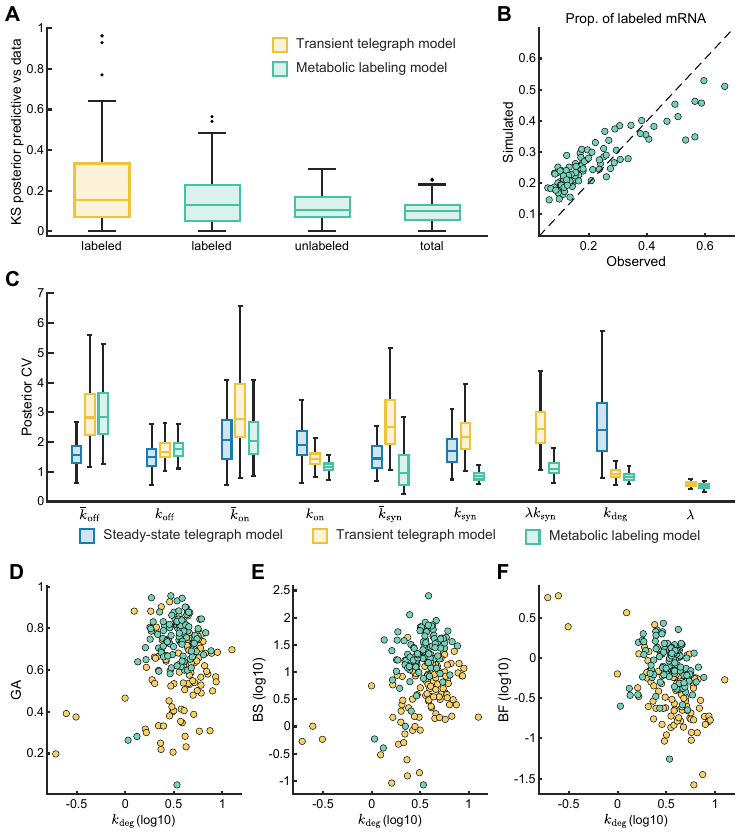}
\caption{\textbf{Validation of Posterior Estimates for Inference of the Transient Telegraph and Metabolic Labeling Models.} \textbf{(A)-(B)} KS metrics between the posterior predictive and the observed data. We compare the labeled counts for the transient telegraph model, and compare labeled, unlabeled, and total counts for the metabolic labeling model. \textbf{(B)} Proportion of labeled mRNA in total mRNA in observed and simulated data. \textbf{(C)} Posterior CVs for inferred kinetic parameters across the 100 genes. The scaled parameters are defined as $\bar{k}_{\text{on}}\coloneqq k_{\text{on}}/k_{\text{deg}}$ and $\bar{k}_{\text{off}}\coloneqq k_{\text{off}}/k_{\text{deg}}$, with $\bar{k}_{\text{syn}}\coloneqq k_{\text{syn}}/k_{\text{deg}}$ for the steady-state telegraph and metabolic labeling models and $\bar{k}_{\text{syn}}\coloneqq \lambda k_{\text{syn}}/k_{\text{deg}}$ for the transient telegraph model. Outlier markers are omitted for visual clarity. \textbf{(D)-(F)} Posterior median of burst kinetics gene activity (GA), burst size (BS), and burst frequency (BF) against that of the degradation rate.}
\end{figure}

\newpage
\begin{figure}
\centering
\includegraphics{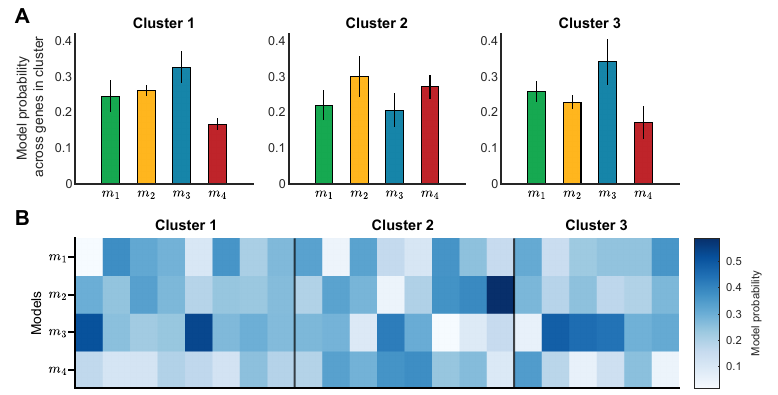}
\caption{\textbf{Model Probabilities Inferred for the Telegraph Model with Cell Cycle-Dependent Rates.} \textbf{(A)} Model probabilities across genes in each cluster (clusters defined as described in the main text). Bars show cluster-level mean model probabilities $\pm$ standard error of the mean across fitted genes. \textbf{(B)} Model probability per fitted gene.}
\end{figure}